\documentclass[fleqn,usenatbib]{mnras}

\usepackage[T1]{fontenc}
\usepackage{ae,aecompl}

\usepackage{graphicx}	
\usepackage{amsmath}	
\usepackage{amssymb}	
\usepackage{pdflscape}	
\usepackage{multicol}
\usepackage{afterpage}
\usepackage{color}
\usepackage{etoolbox}
\usepackage[utf8]{inputenc}
\usepackage{ulem}

\definecolor{darkmagenta}{rgb}{0.5, 0, 0.5}
\definecolor{darkgreen}{rgb}{0, 0.6, 0.05}
\definecolor{darkred}{rgb}{0.86,0.078,0.235}

\title[Central Image]{Constraints on the Inner Regions of Lensing Galaxies from Central Images using a Recent AGN Offset Distribution}

\author[D. Perera et al.]{D. Perera$^{1}$\thanks{E-mail: perer030@umn.edu},
L. L. R. Williams$^{1}$, C. Scarlata$^{1}$
\\
$^{1}$School of Physics and Astronomy, University of Minnesota, Minneapolis, MN, 55455, USA.\\
}
\date{Accepted XXX. Received YYY; in original form ZZZ}

\pubyear{2022}

\begin{document}
\label{firstpage}
\pagerange{\pageref{firstpage}--\pageref{lastpage}}
\maketitle

\begin{abstract}
In gravitational lensing, central images in quads can serve as a powerful probe of the inner regions of lens galaxies. The presence of an offset central supermassive black hole (SMBH) has the potential to distort the time-delay surface in a way such that 3 central images form: a strongly demagnified image near the SMBH, and two less demagnified (and potentially observable) images at a central maximum and saddle point. Using a quad-lens macro-model, we simulate the constraints that could be placed on various lens galaxy parameters based on their central images’ probability of detection or non-detection. Informed by a recent low-redshift distribution of off-nucleus AGN, we utilize Bayesian inference to constrain the mean SMBH off-nucleus distance and galactic core radius for a sample of 6 quads. In general, we find that a detection of the central image in any quad would favor larger SMBH off-nucleus distances and galaxy core sizes. Assuming a linear relationship between core radii and velocity dispersion $r_c = b\sigma$, these results similarly imply strong constraints on $b$, where the likely case of a central image non-detection in each quad constraining $b$ to $3.11^{+2.72}_{-2.26} \times 10^{-4}$ kpc km$^{-1}$ s. Our results show that tight constraints on lens galaxy parameters can be made regardless of a detection or non-detection of a central image. Therefore, we recommend observational searches for the central image, possibly using our suggested novel detection technique in UV filters, to formalize stronger constraints on lens galaxy parameters.

\end{abstract}



\section{Introduction}

Gravitational lensing theory predicts that the number of multiple images formed by non-singular mass distributions must always be odd. Following Fermat's Principle, the locations of these images correspond to stationary points on the time-delay surface. Three image systems (known as "doubles"), form visible images at a minimum and saddle point, where the saddle point image is usually demagnified relative to its minimum counterpart. Five image systems (known as "quads") form 2 images at minima and 2 images at saddle points. In both cases, a demagnified central image is formed at a maximum near the center of the lens. The central image is usually demagnified beyond visibility due to steep central lens density profiles causing sharp peaked behavior of the time-delay surface.  Almost all existing searches for the central image rely on the optical, or radio wavelengths. In the optical, the central image is drowned by the light of the lensing galaxy, while detections in the radio are limited because most QSO sources are radio quiet. 

Out of about 200 known doubles, only 2 have observed central images: PMN J1632-0033, demagnified to 0.004 (by 6 mag.), compared to the brightest image \citep{winn04}, and PKS 1830-211, demagnified to 0.007 
\citep[by 5.4 mag.][]{muller20}. Of the $\sim$50 known quads lensed by an isolated galaxy, no detections exist in the optical, radio, or at ALMA wavelengths \citep{wong15,tamura15,wong17b}. The more recently discovered systems from {\it{ Gaia}} \citep{lemon22} do not appear to contain central images, but that is yet to be confirmed with further observations. 

Even though the central image is seldom detected, its study is of great interest because it can serve as a useful probe of the central regions of lens galaxies and Supermassive Black Hole (SMBH) properties. Most, and probably all galaxies have SMBHs at their centers \citep{kormendy95,ferrarese00}. The black holes are tied to two important aspects of galaxies: SMBH growth is closely linked to galaxy formation, including galaxy mergers \citep{dimatteo05,koss18} and formation of density cores \citep{nasim21}, and inspiralling and mergers of binary SMBH \citep{begelman80} leading to the emission of gravitational waves.

An SMBH at kpc distances from the galaxy center will slowly spiral inwards through dynamical friction from collisionless particles, namely stars and dark matter, as well as gas \citep{chen22}. Observationally determined distribution of galaxy host-SMBH separations will constrain these dynamical friction timescales, and test hydrodynamical cosmological simulations \citep{volonteri20,katz20}. In massive elliptical galaxies, this process will carve out a density core. To get a better physical understanding of the dynamical friction within $\sim$1 kpc of the galaxy center it is crucial to constrain the sizes of galaxies’ density cores. Nearby ($\lesssim$ 100 Mpc) massive galaxies with large ($\sim$700 pc) density cores can be resolved \citep[e.g.][]{rantala18,thomas16}, but this is more difficult for more distant galaxies and smaller cores. Since the central lensed image is affected by the SMBH and the galaxy mass density near the center \citep{mao01,rusin05,mao12}, a detection or upper limit on its brightness can place constraints on SMBH mass, distance from center, and lens galaxy core size. 

\begin{figure*}
    \centering
    \includegraphics[trim={0cm 3cm 0cm 1.0cm},clip,width=\textwidth]{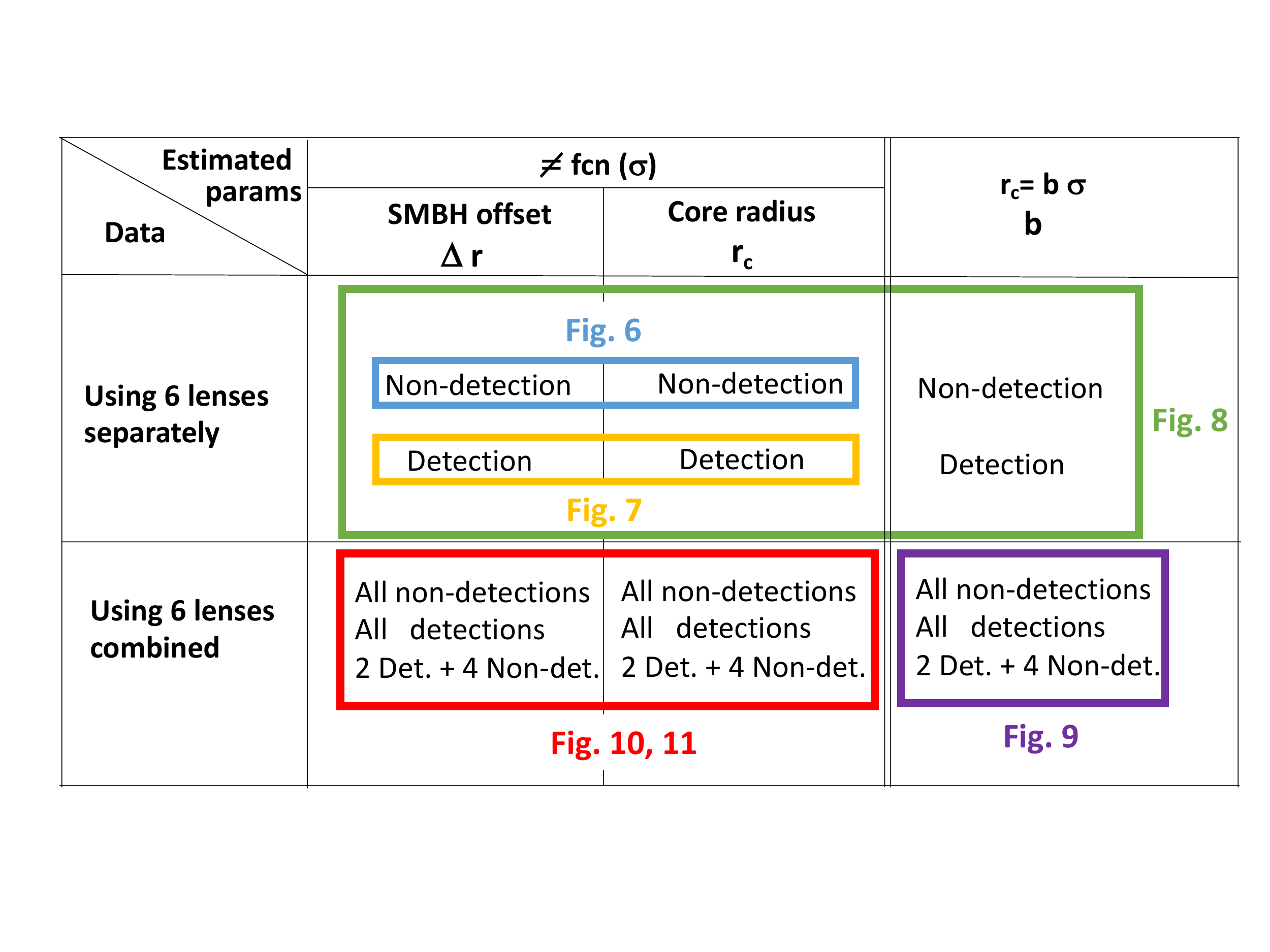}
\caption{Summary of our statistical analysis plan. We consider two assumptions for inference: (1) $r_c \not \propto \sigma$ and (2) $r_c \propto \sigma$. In the first case (1), we constrain the SMBH offset and galaxy core radius for individual quads depending on a hypothetical central image non-detection (blue and Figure \ref{fig:indivnondetcontours}) or detection (yellow and Figure \ref{fig:detcontours}). The cases in the red box, and Figures \ref{fig:allpdfs} and \ref{fig:totcontours} make an assumption that all galaxy lenses have the same SMBH offset and core radius, which is unlikely to be true. Therefore these estimates should be treated only as "back-of-envelope" for this general type of galaxies. In the second case (2), we only make an inference on the $r_c - \sigma$ proportionality constant $b$, which is assumed to be the same for all galaxies. The constraints on $b$ from individual quads are given for each detection scenario (green and Figure \ref{fig:indivpost}). Combining these results for a global constraint on $b$ is similarly given for each detection scenario (purple and Figure \ref{fig:allpdfs_b}).}
\label{fig:analysis}
\end{figure*}

Using lens system CLASS B1030+074, where no central image was detected, \citet{quinn16} finds that $\sim$45\% of galaxies should yield observable central images, assuming demagnifications $\leq$10 magnitudes relative to the brightest image.
\citet{hezaveh15} propose that 10-hr ALMA observations in the $mm$ band can detect a central image at high significance for a lens galaxy core size $\geq$0.2 kpc, allowing for strong constraints on central density slope, core size, and mass of the central SMBH. 
Recent ALMA observations have yielded upper limits on the central image flux and SMBH mass \citep{wong15,tamura15,wong17b}.

If measured or constrained through observations, properties of SMBH, such as their mass and distance from the host galaxy center will provide invaluable clues for the understanding of the central regions of galaxies, and SMBH merger rates. The masses of SMBH in many nearby galaxies have been measured using a range of other methods \citep{dullo21,gultekin09}. However, their distances from host galaxy centers are less well known \citep{skipper18}.

A recent examination of $z=0.3-0.8$ Active Galactic Nuclei (AGN) led to a determination of the characteristic probability density function of their offsets' upper limits from the host galaxy center \citep{shen19}. Their results and sample are given the name of VODKA, which we accordingly adopt. The employed methodology, known as "varstrometry", is described lucidly in \citet{hwang20}. (This technique can be applied to single off-nucleus AGN, or generalized to dual AGN.)

Observationally, a sub-kpc off-nucleus AGN and the center of its host galaxy appear as single-source photocenter (photometric center) for $z>0.5$ \citep{hwang20}. The vast majority of AGN exhibit aperiodic photometric variability on day-year timescales of $\gtrsim 0.03$ mag \citep{sesar07}, with $\sim$30\% varying $\geq$0.1 mag \citep{sesar07,rengstorf06}. For a single off-nucleus variable AGN and constant-flux host galaxy, it is expected that the AGN variability will lead to astrometric variability of the photocenter of the AGN-galaxy system, which will be strongly correlated with the total detected flux \citep{hwang20}. This allows for a measurement of the distance separation between the AGN and host galaxy center with linear regression \citep[see eq. 4 of ][]{hwang20}. When applied to low redshift ($0.3 < z < 0.8$) AGN and host galaxy pairs observed with {\it Gaia} DR2, it is found that there are strong constraints on AGN separation. Nearly all AGN are at $<$ 1 kpc, 90\% are at $<$ 500 pc, and 40\% are at $<$ 100 pc \citep{shen19}. While this result is a significant improvement over previous determinations, it is important to realize that the distribution of separations of AGN may differ from that of SMBH. All SMBH, not just AGN affect the central regions of galaxies by reshaping their central mass distributions, and also lead to the formation of gravitational wave emitting inspirals. 

Offset SMBH are of particular importance for studying and potentially detecting the central lensed image. Ideally, one wants the distribution of host galaxy-SMBH separations, but such information does not exist. Instead, we take the AGN-galaxy host separation distribution from VODKA and use it as a prior in our analysis. The present paper is the first to incorporate the distribution of AGN offsets.

\begin{figure*}
    \centering
    \includegraphics[trim={0cm 0cm 0cm 1.0cm},clip,width=0.49\textwidth]{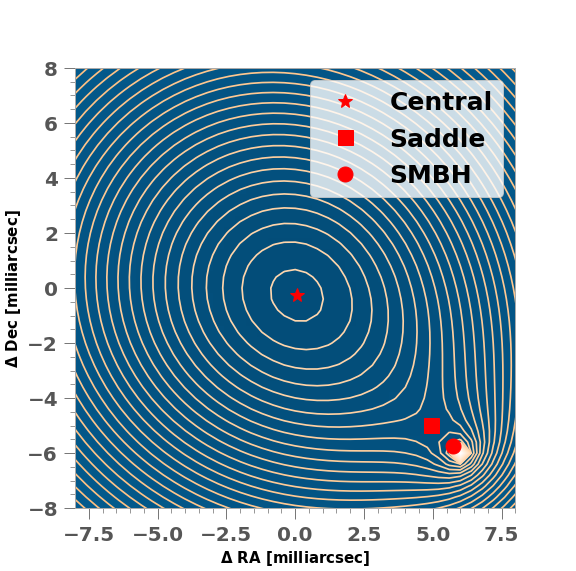}
    \includegraphics[trim={0cm 0cm 0cm 2.0cm},clip,width=0.49\textwidth]{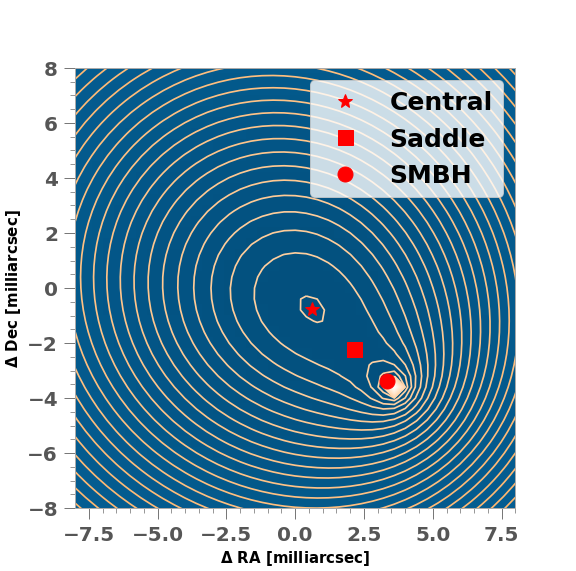}
\caption{Visualization of the Fermat lens potential distortion by the SMBH. In both panels, the view is restricted to the central 256 mas$^2$ of the lens galaxy such that the 4 quad images are out of view. The {\it left panel} shows the resulting Fermat potential contours when a $3 \times 10^8 M_{\odot}$ SMBH is introduced at $(\Delta {\rm RA},\Delta {\rm Dec}) = (6.0,-6.0)$ mas. In this case 3 images form (for a total of 7 in the system): a central image at the central maximum, a saddle image between the central maximum and the SMBH, and a SMBH image very close to the SMBH maximum. The spacing of contours shows the steepness of the potential increase due to SMBH, indicating that the SMBH image is strongly demagnified. The {\it right panel} shows the lens potential contours when the same SMBH is offset a shorter distance at $(\Delta {\rm RA},\Delta {\rm Dec}) = (3.6,-3.6)$ mas. The same 3 images form, with the saddle images forming closer to the midpoint of the central and SMBH images due to the less offset SMBH. The images in this case (right panel) are all magnified relative to the images in the first case (left panel), illustrating the dependence of central image magnification $\mu$ on SMBH offset $\Delta r$.
}
\label{fig:lpot}
\end{figure*}

Gravitational lensing theory predicts that a sufficiently offset SMBH can produce extra central images. This is displayed in Figure \ref{fig:lpot}. The inclusion of an offset SMBH creates two additional stationary points (and thus two new images) in the time-delay surface: a steep maximum very near to the location of the SMBH and a saddle between the original central maximum and the offset SMBH. The formation of the two new images and their properties depend on the offset distance of the central SMBH, the azimuthal location of the SMBH in the case of non-circular lenses, and the lens galaxy core size. If the offset distance is too short, then only one image forms at the SMBH maximum. The SMBH (maximum) image is always strongly demagnified, regardless of whether the two other central images are formed. When additional central images form in the case of a large offset distance, the central image near the maximum of the lens density profile, and the saddle image are not as demagnified, and can potentially be bright enough to be observed. It is important to emphasize that if the SMBH is not offset sufficiently, then no extra central images will form.

Here we present a new modelling framework that will allow constraints to be made on various galaxy properties with future observations. To make predictions about these properties, we proceed in three sequential steps. In the first step we create the macro-model of the galaxy using the positions and time delays of the four images of its quad. In the second step, we simulate central images for each of the macro-models. We treat the SMBH as a point mass and sample its offset according to the aforementioned VODKA distribution. Lastly, we use Bayesian inference to simulate constraints on galaxy parameters for specific observation scenarios.

Our analysis is the first one to study several (6) quad lens systems. We restrict our sample to quads because they provide more constraints for the galaxy macro-model, which affects the central region through its ellipticity and the location of the source.  We assume two main possible scenarios for each quad: 
(i) a non-detection of central images 
(ii) detection of at least one of the central images.
This allows us to place statistical constraints on the SMBH offset ($\Delta r$) and core radius ($r_c$) of individual lenses. 

Additionally, we want to constrain the global distribution of these parameters for our galaxies. However, the distributions of core radii and SMBH distances may depend on the galaxy's mass, or velocity dispersion $\sigma$ (and probably other parameters), and will not be the same for all galaxies in our sample. If one wants a ballpark value of a parameter of typical lens galaxy, one can ignore these differences, and estimate the parameter by combining the results for individual quads. We also consider a separate scenario of combination of detection in some, and non-detection in other quads in our sample. This analysis assumes $r_c \not \propto \sigma$. Alternatively, if we assume that $r_c=b\sigma$, then $b$ is the same for all galaxies, and can be estimated using all quads combined. This analysis plan is outlined in Figure \ref{fig:analysis}. 

Importantly, each scenario yields independent statistical constraints on the SMBH offset distribution and central density profile based on prior distributions informed from the VODKA analysis and our lens macro-model sample. 
Assuming that lenses from ongoing and future surveys will be analogous to the sample we use, our analysis can be used to forecast how well lens galaxy properties can be constrained with future data. 

In Section \ref{txt:data} we describe our sample of lensed QSOs; in Section \ref{txt:modelling} we discuss our lens macro-modelling, simulation of the central images, and statistical inference analysis; in Section \ref{txt:results} we present the results of our analysis; in Section \ref{txt:detection} we describe a novel technique to detect central images; and in Section \ref{txt:conclusions} we discuss the implications of our constraints.

\section{Data}\label{txt:data}

\begin{table*}
	\caption{Sample of gravitationally lensed QSOs}
	\begin{tabular}{llllllllc} 
		\hline
		QSO & z$_{{\rm QSO}}$ & z$_{{\rm lens}}$ & $\sigma$ [km s$^{-1}$] & M$_{\rm SMBH}$ [M$_{\odot}$] & m$_{o}$ [AB], Filter & Flux [$10^{-17} f_{\lambda}$]  & Time Delays & References\\
		\hline
		HE0435-1223 & 1.693 & 0.454 & 222 $\pm$ 15 & 3.7 $\pm$ 1.2 $\times 10^8$ & 20.33, F275W & 10.36 $\pm$ 0.07 & $\Delta\tau_{12}, \Delta\tau_{13}, \Delta\tau_{14}$ & 1, 8 \\
		PG1115+080 & 1.735 & 0.311 & 287 $\pm$ 18 & 1.3 $\pm$ 0.4 $\times 10^9$ & 18.60, F218W & 101.30 $\pm$ 0.50 & $\Delta\tau_{14}, \Delta\tau_{\{23\}4}, \Delta\tau_{1\{23\}}$ & 2, 8, 10\\
		RXJ1131-1231  & 0.657 & 0.295 & 323 $\pm$ 20 & 2.3 $\pm$ 0.7 $\times 10^9$ & 19.62, F218W & 32.96 $\pm$ 0.29 & $\Delta\tau_{14}, \Delta\tau_{24}, \Delta\tau_{23}$ &
		3, 8\\
		SDSSJ0924+0219 & 1.523 & 0.393 & 215 $\pm$ 21.5 & 3.2 $\pm$ 1.6 $\times 10^8$ & 19.41, F275W & 25.40 $\pm$ 0.12 & --- & 4\\
		WFI2033-4723 & 1.662 & 0.658 & 250 $\pm$ $^{15}_{21}$ & 6.7 $\pm$ 2.3 $\times 10^8$ & 18.56, F467M & 18.27 $\pm$ 0.23 & $\Delta\tau_{1\{23\}}, \Delta\tau_{14}, \Delta\tau_{\{23\}4}$ & 5, 8, 11\\
		MG0414+0534 & 2.640 & 0.958 & N/A & 8.0 $\times 10^{8\pm1}$ & 23.58, F621M & 0.11 $\pm$ 0.01 & --- & 6, See Section \ref{txt:data}\\
		RXJ0911+0551 & 2.763 & 0.769 & N/A & 8.0 $\times 10^{8\pm1}$ & 19.45, F547M & 6.11 $\pm$ 0.10 & $\Delta\tau_{1\{234\}}$ & 7, 9, See Section \ref{txt:data}\\
		\hline
	\end{tabular}\\
\medskip{The columns list QSO identifier, source and lens redshifts, lens velocity dispersion, SMBH mass inferred from the $M-\sigma$ relation, estimated apparent magnitude m$_{o}$ for the brightest quad image as observed in the corresponding HST filter, specific flux of the brightest quad image in the same HST filter ($f_{\lambda}$ in units of ergs s$^{-1}$ cm$^{-2}$ ~\AA$^{-1}$), the time delays used, and references from which the data were taken. The subscripts on the time delays $\Delta\tau$ indicate the arrival order of the two images. Curly brackets enclose images assumed to have the same arrive time. Two systems did not have published image arrival times.}
\medskip{References: (1) \citet{courbin11,wong17}, (2) \citet{tonry98,treu02}, (3) \citet{sluse07,suyu13}, (4) \citet{macleod15}, (5) \citet{sluse19}, (6) \citet{hewitt92}, (7) \citet{bade97}, (8) \citet{millon20}, (9) \citet{hjorth02}, (10) \citet{bonvin18}, (11) \citet{bonvin19}
}
\label{tab:qso_sample}
\end{table*}

Table~\ref{tab:qso_sample} presents our sample of gravitationally lensed QSOs. This sample was chosen to roughly span the redshift range of the VODKA population, $0.3\!<\!z\!<\!0.8$ \citep{shen19}, allowing direct comparison of our results with their AGN offset distribution. Most of our systems also have time delay information, which aids in lens macro-modeling. The SMBH masses we use are found by applying the $M-\sigma$ relation from \citet{dullo21}, as explained in further detail in Section \ref{txt:imagegeneration}. Four of the lens galaxies have measured central velocity dispersions. For SDSSJ0924+0219, \citet{macleod15} does not directly measure $\sigma$, and instead estimates it with lens modelling. We assume 10\% uncertainty on their result, and record that value in Table~\ref{tab:qso_sample}. MG0414+0534 and RXJ0911+0551 have no measured $\sigma$ in the literature. Therefore, we assume a $\sigma$ for these two systems equal to the average of the measured $\sigma$ in the rest of the sample. From this we obtain their $M_{\rm SMBH} \approx 8.0 \times 10^8 M_{\odot}$ from the same $M-\sigma$ relation. For the uncertainty on this $M_{\rm SMBH}$, we assume a factor of 10, so as to span a wide enough range for possible $M_{\rm SMBH}$. 

RXJ0911+0551 presents itself as an exception in our sample. The radial distribution of the 4 observed quad images is highly asymmetric, probably due to the presence of a nearby galaxy cluster in the field, providing external shear \citep{burud98,tortora04}. As a result, the core size distribution of our lens macro-models skews to larger values beyond the range of the other 6 lenses in our sample. Therefore, we treat this lens separately from the rest of the sample of 6 and derive independent constraints from it (see Table \ref{tab:indivconstraints} and Section \ref{txt:rxj0911}). Henceforth, we refer to "our sample" as those QSOs mentioned in Table \ref{tab:qso_sample} not including RXJ0911+0551.

\section{Analysis}\label{txt:modelling}
The goal of the analysis is to obtain constraints on the galaxy core radii and SMBH offset using observational constraints on the central image.

Our analysis can be summarized in three sequential distinct steps: (1) We constrain  galaxy macro-models based on the 4 non-central images. The result of this step are $1000$ macro-models per lens, each described by 16 parameters, including the QSO source $(x,y)$ position. The lens galaxy core radius is derived from the 14 lens parameters. These macro-models form a prior for the next step of the analysis. (2) For each macro-model, we simulate the central, saddle, and SMBH image locations and magnifications using a large sample of SMBH offsets from the VODKA distribution and SMBH masses from the $M-\sigma$ relation. Image properties depend primarily on the SMBH masses ($M_{\rm SMBH}$), offsets ($\Delta r$), and galaxy's core radius ($r_c$). (3) The final step is the statistical inference for the two main observational scenarios we consider: (i) non-detections of the central image, and (ii) detection of at least one central image. Additionally, when we combine these constraints for all galaxies, we also consider a third case of some detections and some non-detections. With this framework, we can write the posterior probability as:
\begin{equation}
\begin{split}
P(\Delta r, r_c | D) \propto \int P(D | M_{\rm SMBH}, \Delta r, r_c)\,P(\Delta r)\,P(r_c) \\ \times\, P(M_{\rm SMBH})\,dM_{\rm SMBH}\label{eq:post1}
\end{split}
\end{equation}
where $D$ are the input image positions and time delays (where available; see Table~\ref{tab:qso_sample}). 

As we explain later in this section, the 16 parameter priors going into the macro-modeling in step (1) are flat. The resulting distributions of these parameters emerging from this step are not flat anymore, as the galaxy properties have been constrained by the quad images. These distributions become the priors for step (2), from which we obtain the prior for the galaxy core radius $P(r_c)$. The distributions of galaxy parameters ($\Delta r$ and $r_c$) get further constrained as a result of step (3), and the constraints are different for non-detections vs. detections. The priors on SMBH properties going into step (2) are described in more detail in Section~\ref{txt:imagegeneration}. Step (3) further constrains these properties, which we present as the main result of the paper.

\subsection{Constraining Galaxy Macro-Model}\label{txt:modelgeneration}

We generate galaxy-scale macro-models based on the 7 lens systems we consider in this paper. Since the central images are currently not observed in any of these, and we envision that our analysis can be extended to future lens systems, our modelling need not be tailored exactly to these systems. 
Instead, we use these 7 as approximate examples of realistic lenses. Because of that, the many macro-models we generate per lens fit the observed lensing data (image positions and time delays, where available) only to $\chi^2\leq 9$, and in the systems where satellite galaxies are detected in addition to the main lens galaxy, their positions are not fixed at the observed position.
Loosening the criteria to accept macro-models with $\chi^2\leq 9$ results in lens plane image rms between $<0.005"$ and $\sim 0.035"$.

We represent lens galaxies by a superposition of two softened power-law ellipsoid potentials, called {\tt alphapot} \citep{kee11}:
\begin{equation}
    \Psi_{\rm gal} = b\left(s^2 + x^2 + \frac{y^2}{q^2} + K^2xy\right)^{\frac{\alpha}{2}}
\end{equation}
where $b$ is the normalization, $\alpha$ is the power-law exponent\footnote{Not to be confused with the deflection angle.}, $s$ is the core radius, $K$ and $q$ determine the ellipticity and the position angle of the ellipsoid \citep{gho20,bar21}. In addition to being a reasonably good representation of elliptical galaxies, it allows for analytical calculations of the deflection angles, normalized projected surface density $\kappa$, and shear $\gamma$ from the first and second derivatives of $\Psi_{\rm gal}$. We do not include the SMBH at this point because it does not affect the QSO image positions, and therefore does not affect the lens galaxy macro-model.

The two mass components were allowed to have a non-zero offset between their centers, $(x_{\rm off},y_{\rm off})$. The reason for the offset is that a single component, or two co-centered components sometimes cannot reproduce QSO images to astrometric precision or other observables of the system \citep{rus20}. Offsets break the elliptical symmetry of the lens and result in lopsided galaxy mass distributions, which apparently help to model some lenses and populations of lenses \citep{bru16,gom18,nig19,wil20,bar21}. 

The two {\tt alphapots} have a total of 10 parameters, i.e., two sets of $b$, $\alpha$, $s$, $q$, and $K$. Combined with $(x_{\rm off},y_{\rm off})$, external shear amplitude and direction, and the QSO source position the total number of macro-model parameters is 16. We use downhill simplex to find solutions. The starting ranges for the macro-model parameters are the same for all 7 lenses. The density slopes $\alpha=1.0\pm 0.2$, core radii $s=200\pm 199$pc, external shear amplitude, $\gamma=0.1\pm 0.1$, and offsets of the secondary mass component, $x_{\rm off}=y_{\rm off}=0\pm 40$pc. We made exceptions for 3 systems that have visible nearby satellite galaxies: WFI 2033, MG 0414, and RXJ 0911, where the second mass components were given larger initial offsets to represent the satellite. For RXJ 0911+0551 we used a larger initial external shear, $\gamma=0.1+0.1\pm 0.1$, to account for the nearby galaxy cluster.

Downhill simplex search is free to modify these initial values, so many of the final values of all parameters were outside of the starting ranges. This is also true of the location of the satellite galaxy; we did not fix it at the observed position but allowed simplex to find different solutions for each run. Because the number of macro-model parameters exceeds the data constraints, we generate many models for each observed quad. From these we reject macro-models that have two density peaks, and those where the single density peak is not coincident with the light peak, i.e., the center of the main lens. We also restrict the ellipticity of the final macro-models to disallow very elliptical or unphysically shaped galaxies: the ellipticity of the lens potential of each of the two mass components were restricted to have axis ratios $\geq 0.7$.  The surviving macro-models---about 1000 per lens system---sample the model space allowed by our assumptions and lensing degeneracies. 

For each lens, the fitted macro-models have a range of central density profile slopes, and hence core radii. In general, steeper central density profiles imply smaller core radii. For each lens macro-model we calculate the galaxy density core size as the radius where the log-log density becomes steeper than $-0.425$. While this value is somewhat arbitrary, it works well to estimate core sizes of mass distributions that do not have a core radius explicitly incorporated in their analytical form, as is the case with our two-component macro-models. Since we generate a range of macro-models for each observed lens, we also have a range of core sizes. Six of the 7 systems have core radii distributions that peak at $\lesssim 100$ pc. RXJ 0911 is an exception: its distribution is broad with a peak at $\sim 400$ pc. These value are consistent with the range determined based on local ellipticals, 50-500 pc \citep{fer06,hezaveh15}.

The 16 parameters for every one of the 1000 macro-models per system are passed to the next step of our analysis. These parameters do not have flat distributions, as they have been constrained by the quad images. Galaxy core size is calculated based on these parameters; its distribution is shown as yellow curves in the Figures \ref{fig:allpdfs} (for the 6 in our sample) and \ref{fig:RXJ0911pdfs} (for RXJ0911) of this paper.

\subsection{Simulating Central Images}\label{txt:imagegeneration}
With the galaxy macro-models generated, the next focus is to find the locations and magnifications of the corresponding central images. Since we are most concerned with the central region, we re-scale the galaxy window to 0.3"$\times$0.3" about the center of the lens.
We generate lensing potentials $\Psi_{\rm gal}$ from the galaxy macro-models parameters (see Section~\ref{txt:modelgeneration}) and SMBH parameters. 

For the SMBH, we assume it to be a point mass with lens potential:
\begin{equation}
    \Psi_{\rm SMBH} = \theta_E^2 \ln{\sqrt{x^2 + y^2}},
\end{equation}
where $\theta_E$ is the Einstein radius of the SMBH:
\begin{align}
    \theta_E &= \sqrt{\frac{4GM_{\rm SMBH}}{c^2}\frac{D_{ds}}{D_sD_d}}.
\end{align}
As mentioned earlier, the perturbation of the outer 4 images in each system is negligible since the SMBH only adds $\sim$10$^{-3}$ times the mass within the Einstein radius to the galaxy. From this setup we simply sum the two lens potential components to get the total lensing potential of a particular system: $\Psi_{\rm tot} = \Psi_{\rm gal} + \Psi_{\rm SMBH}$. We then scan the re-scaled window to find the locations and magnifications of the images.

Within this framework, we displace the SMBH from the galaxy center according to the results from VODKA. To do this we generate a large sample of SMBH offset positions according to the probability distribution obtained by VODKA. From the right panel of Figure 1 of \citet{shen19}, we transform the presented CDF into a PDF and fit to a Gaussian profile to obtain the best fit average and standard deviation of their sample. We find $0.131\pm 0.008$ kpc, and $0.163\pm 0.007$ kpc for the average and standard deviation, respectively. In practice, this represents a truncated Gaussian since the distribution extends into the regime of negative SMBH offsets. Using a Box-Muller transform, we use these results to generate a distribution of SMBH-galaxy center offsets. The azimuthal locations of SMBH with respect to the galaxy center were picked randomly. It is important to note that VODKA provides the distribution of SMBH offset upper limits, and we assume that the offset positions are the same as these upper limits.
This is supported by recent simulation results which find that 60\% of SMBH in brightest cluster galaxies at $z=0$ are offset by $>0.1\,$kpc, and at $z=2$ about 80\% are offset by $>1\,$kpc \citep{chu22}. While the lens galaxies in our sample are not BCGs, their stellar masses are comparable to those in that study.

Using the $M-\sigma$ relation for Sersic plus Core-Sersic galaxies from \citet{dullo21}, and the velocity dispersion, $\sigma$ (Table~\ref{tab:qso_sample}) for each QSO lens galaxy, we obtain a SMBH mass range for each lens. This range is determined by propagating the measured uncertainty in $\sigma$ in the $M-\sigma$ relation. For each macro-model generated, we pick the SMBH mass randomly from within this range.

Finally, for each of the 1000 macro-models per system we generate images based on the following parameters described above: SMBH mass, SMBH offset distance, SMBH azimuthal position, and lens galaxy macro-model (each with a corresponding galaxy core size). 
In general, the central image is always produced and is always demagnified. However, given certain SMBH masses, offsets, and macro-models, two additional images can be produced: a similarly demagnified saddle image, and a more strongly demagnified image very near the SMBH (see Figure~\ref{fig:lpot}). The saddle image always forms between the central and SMBH image. Using the results of magnification and SMBH offset for each macro-model, we can place statistical constraints on the galaxy lens core size and SMBH offset.

\subsection{Statistical Inference}\label{txt:statistics}

In this section we model the analysis one would do in case of the likely scenario of a non-detection of the central image. Additionally, we also consider a less likely but more interesting case of a central image detection. Deep observations, whether or not they result in a central image detection, can constrain the galaxy core size and SMBH offset.

To place constraints on the galaxy core size $r_c$, and SMBH offset $\Delta r$, we employ Bayesian inference based on the results from our analysis described in Sections \ref{txt:modelgeneration} and \ref{txt:imagegeneration}. Our first analysis assumes that $r_c$ is independent of the galaxy's measured velocity dispersion $\sigma$. Later, when we combine all the quads in our sample for a general constraint, we assume that $r_c$ is proportional to $\sigma$, which is the simplest, yet physically plausible, relation that these two parameters can have. For each of these two analyses we consider two cases to inform our likelihood function: non-detection of all central images and detection of at least one of the central images, for example, using the technique outlined in Section \ref{txt:detection}. In the latter analysis, we consider a case where 2 systems have detections, and 4 have non-detections.

Focusing first on results for individual QSO quads with no $r_c - \sigma$ relation, we assume truncated Gaussian prior probabilities on SMBH offset $\Delta r$ according to the VODKA distribution \citep{shen19}, and prior probability distribution of the galaxy core sizes, $r_c$, from the galaxy lens macro-models, which resemble Gaussians, described in Section~\ref{txt:modelgeneration}. For the SMBH mass $M_{\rm SMBH}$ we assume a uniform prior $P(M_{\rm SMBH})$ within the defined $M_{\rm SMBH}$ range described in Section \ref{txt:imagegeneration}. With this, we can solve for the 2D posterior probability using equation \ref{eq:post1}.

To get a posterior on $r_c$ for each quad, we can marginalize $P(\Delta r, r_c | D)$ over $\Delta r$: 
\begin{equation}
\begin{split}
P(r_c | D) \propto \int_{\Delta r} P(\Delta r, r_c | D) d\Delta r \label{eq:postrc}
\end{split}
\end{equation}
We can repeat vice versa for a posterior on $\Delta r$. This intrinsically assumes that there is no relation between $r_c$ and $\sigma$. We apply this analysis individually to each quad to estimate their $\Delta r$ and $r_c$.

Alternatively, we can expand the above method to account for a potential correlation between velocity dispersion and core size. As stated above, we assume that $r_c$ and $\sigma$ are proportional, therefore:
\begin{equation}
    r_c = b\sigma \label{eq:rsigma}
\end{equation}
where $b$ is a fit parameter we wish to extract. Under this model, the parameter $b$ is assumed to be the same for all quads, and we do not attempt to constrain $\Delta r$. From this, we assume priors for $b$, $P(b)$, with the same shape as $P(r_c)$. We continue to use the same priors for $M_{\rm SMBH}$ and $\Delta r$ as before. Therefore, we can write a separate 2D posterior:
\begin{equation}
\begin{split}
P(\Delta r, b | D) \propto \int P(D | M_{\rm SMBH}, \Delta r, b)\,P(\Delta r)\,P(b) \\ \times\, P(M_{\rm SMBH})\,dM_{\rm SMBH}\label{eq:altpost1}
\end{split}
\end{equation}
Repeating the same marginalization as in equation \ref{eq:postrc} will give us individual posteriors for each lens for $b$. To get a global constraint on $b$, we can multiply all the resulting marginal posteriors for each quad together:

\begin{equation}
\begin{split}
P(b | D) \propto \prod_{\rm quads} \left( \int_{\Delta r}  P(\Delta r, b | D)  d\Delta r \right) \label{eq:post4}
\end{split}
\end{equation}
This allows us to constrain the $r_c - \sigma$ relation.

We emphasize that the constraints for $\Delta r$ and $r_c$ are results applicable to individual lenses, thus providing a robust method to derive these properties for observed quads. For $b$, however, the constraint is intended to be a single result for all lenses assuming that equation \ref{eq:rsigma} is obeyed. 

Combining the distributions for $\Delta r$ and $r_c$ presents a separate challenge. As a ballpark estimate, we can use the simplistic assumption that $\Delta r$ and $r_c$ are the same for all lenses. With this, we can derive these constraints in the following way:

\begin{equation}
\begin{split}
P(\Delta r | D) \propto \int_{r_c} \left(\prod_{\rm quads} P(\Delta r, r_c | D) \right) dr_c \label{eq:post3}
\end{split}
\end{equation}
and:
\begin{equation}
\begin{split}
P(r_c | D) \propto \int_{\Delta r} \left(\prod_{\rm quads} P(\Delta r, r_c | D) \right) d\Delta r \label{eq:combinepostrc}
\end{split}
\end{equation}

For the scenarios of a central image non-detection and detection, we follow the outlined inference procedure and simply define different likelihood functions $P(D | M_{\rm SMBH}, \Delta r, r_c)$ for each case, as explained below.

\subsubsection{Central Image Non-Detection}
For the likely case of a non-detection, we define a limiting specific flux $f_{\rm crit}$ below which we assume any central image regardless of its position in the lens plane will not be detected. While this flux limit can be varied at one's wish, we set it to be equivalent to 10 magnitudes fainter than the brightest quad image in each lens system, corresponding to a magnification\footnote{Magnifications can be converted into magnitudes using:
\begin{align*}
    m_i &= -2.5\log_{10}(\mu_i) + m_{\rm BI}
\end{align*}
where $m_i$, $\mu_i$, and $m_{\rm BI}$ are the image magnitude, image magnification, and brightest quad image magnitude, respectively.} $\mu_{\rm crit}$ of 10$^{-4}$ relative to the brightest quad image \citep[same condition as used in][]{quinn16}. Since, by definition, magnification $\mu = f/f_{\rm BI}$ where $f_{\rm BI}$ is the flux of the brightest quad image (see Table \ref{tab:qso_sample}), the limiting flux $f_{\rm crit} = \mu_{\rm crit}f_{\rm BI}$. Under this assumption, we define the likelihood of non-detection $P(D | M_{\rm SMBH}, \Delta r, r_c)$ as the integral of a Gaussian probability density centered around each predicted image flux ($f_i$) up to $f_{\rm crit}$:
\begin{equation}
    P(D | M_{\rm SMBH}, \Delta r, r_c) \propto \int_{-\infty}^{f_{\rm crit}} \exp(-\frac{1}{2}\left(\frac{f - f_i}{\sigma_f}\right)^2) df \label{eq:nondetection}
\end{equation}
The dependence of the likelihood on $M_{\rm SMBH}$ enters through $f_i$. We define the image flux uncertainty $\sigma_f$ to be $0.4 f_i$ to allow for a range of uncertainty in the image flux. Furthermore, we note that the predicted image positions $\vec r_i$ are absent in this likelihood as this information would be unknown in the case of a central image non-detection.

\subsubsection{Central Image Detection}
For the case of a detection of the central image, we define a hypothetical observed central image with specific flux $f_o$ and position in the lens plane $\vec r_o$. For each image formed for a given $M_{\rm SMBH}$, $\Delta r$, and $r_c$, the likelihood of detection is assumed to be Gaussian relative to the hypothetical observation: 
\begin{equation}
    P(D | M_{\rm SMBH}, \Delta r, r_c) \propto \exp(-\frac{1}{2}\left(\frac{(f_i-f_o)^2}{\sigma_f^2}+\frac{(\vec r_i-\vec r_o)^2}{\sigma_r^2}\right)) \label{eq:detection}
\end{equation}
Here, $f_i$ and $\vec r_i$ are the predicted image flux and position for a given $M_{\rm SMBH}$, $\Delta r$, and $r_c$, while $\sigma_f$ and $\sigma_r$ are the uncertainties on $f_o$ and $\vec r_o$. From our method, this requires us to choose the location and flux of detection in each system's lens plane. This can theoretically be anywhere in the lens plane; however, for simplicity, we choose the predicted central image location $\vec r_o$ to be at the center $(0,0)$ of each galaxy lens macro-model. We set $\sigma_r$ based on the resolution of an image of this detection. Assuming an HST image, we adopt the astrometric precision of $\sim$0.03". We convert this $\sigma_r$ into kpc units with the measured $D_d$ for each lens. It turns out that this choice for $\sigma_r$ is relatively large, therefore minimizing the importance of the positional dependence of the likelihood of detection in equation \ref{eq:detection}. However, since a hypothetical detection of the central image depends more on brightness rather than position\footnote{This assumption is valid only because our detection technique (Section \ref{txt:detection}) assumes the lens galaxy is invisible in the observing filters.}, and our choice of $\sigma_f$ is more restrictive, we justify this choice as realistic. For $f_o$ we simply choose the equivalent of 9 magnitudes fainter than the estimated brightest quad image for each lens system. This choice is consistent with our choice of 10 magnitudes fainter for non-detections. We set $\sigma_f = 0.4 f_o$, equivalent to $\sim$1 magnitude in uncertainty. This allows us to encompass a wide range of produced image magnitudes and to account for magnification dispersion from stellar microlensing \citep{dob07}.

Using this likelihood function in equation \ref{eq:post1} and marginalizing over $M_{\rm SMBH}$ gives us the posterior probability density function for $\Delta r$ and $r_c$. 

\section{Results}\label{txt:results}

\subsection{Lens Macro-model Maps}\label{txt:maps}

\begin{figure*}
\begin{multicols}{2}
    \includegraphics[trim={4.0cm 0.35cm 4.5cm 0.37cm},clip,width=0.49\textwidth]{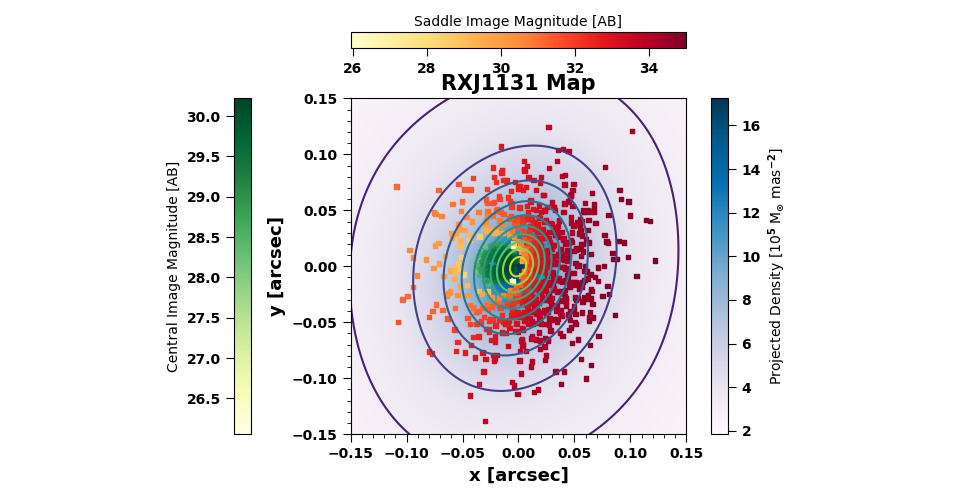}\par
    \includegraphics[trim={4.0cm 0.35cm 4.5cm 0.37cm},clip,width=0.49\textwidth]{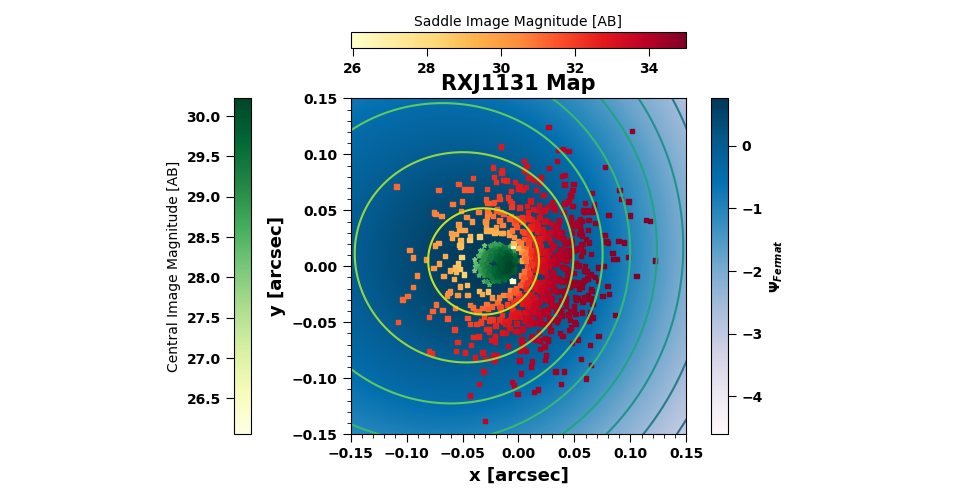}\par
\end{multicols}
\caption{Central image modelling for RXJ1131-1231 quad lens. We used 1576 SMBHs (not shown) randomly offset from the galaxy host center according to the distribution in \citet{shen19}. In this figure all SMBH have $M_{\rm SMBH}\approx 2.11\times10^{9}M_\odot$. (Other maps for this system use different $M_{\rm SMBH}$ randomly chosen from its range shown in Table \ref{tab:qso_sample}).  Offset SMBHs that produce central images (green stars), and saddle images (inverted parity; orange and red squares) are plotted on the mass density ({\it left panel}), and Fermat potential $\Psi_{\rm Fermat}$ ({\it right panel}) of one model of RXJ1131-1231 ($\Psi_{\rm SMBH}$ is not included so as to avoid cluttering the plot). Cases where only one central image is produced at the location of the offset SMBH are excluded from this plot, but included in subsequent calculations. The left and top colorbars indicate the magnitude of the central and saddle images, respectively. The maps have scales of 0.0035 kpc pix$^{-1}$. All our maps have 1250 pix arcsec$^{-1}$. The asymmetry in the image distribution visible in both the panels comes from modeling of the four quad images.}
\label{fig:RXJ1131all}
\end{figure*}

\begin{figure}
    \centering
    \includegraphics[trim={0.0cm 0.35cm 0.0cm 0.0cm},clip,width=0.49\textwidth]{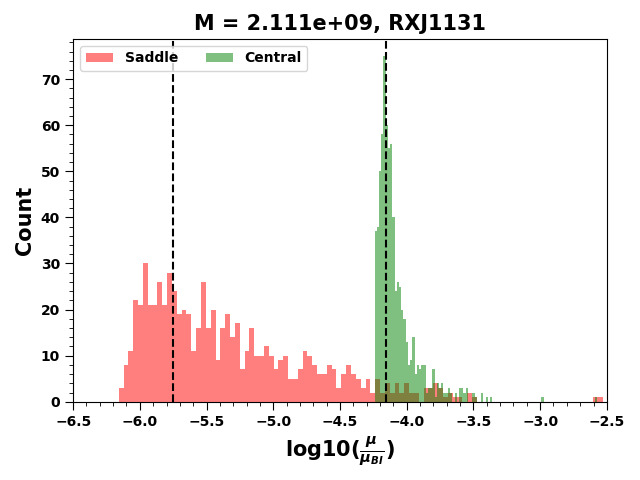}
\caption{Histogram of relative magnifications in log space of central and saddle images for a SMBH of $M_{\rm SMBH}\approx 2.11\times10^{9}M_\odot$ for the case of RXJ1131-1231 shown in Figure~\ref{fig:RXJ1131all}. $\mu$ and $\mu_{\rm BI}$ are the magnifications of the modeled central or saddle image, and the brightest image in RXJ1131-1231, respectively. The two vertical dashed lines correspond to image magnitudes of $m_{\rm AB}=30$ and $34$.
}
\label{fig:RXJhistogram}
\end{figure}

In this section we present examples of maps for some individual galaxy macro-models (see Section \ref{txt:modelgeneration}), and properties of the corresponding central images (see Section \ref{txt:imagegeneration}), before obtaining statistical results in Section~\ref{txt:constraints}.

Following the modelling procedure outlined in Section \ref{txt:modelling}, we obtain maps for each lens galaxy macro-model depicting the locations of SMBH that allow extra central images to form. For illustrative purposes, the left panel of Figure~\ref{fig:RXJ1131all} shows contours of one of these mass maps computed for the case for RXJ1131-1231 with a $2.11 \times 10^9 M_{\odot}$ SMBH\footnote{As shown in Table \ref{tab:qso_sample}, this SMBH mass is roughly near the center of the $M_{\rm SMBH}$ range found from the $M-\sigma$ relation.} displaced according to VODKA distribution, and with random azimuthal positions. The central and saddle images are presented in these maps, as greenish and reddish points, respectively. The SMBH image is typically extremely demagnified, far below observability, so these are not included in the map. Similarly, to avoid cluttering the map, the SMBH locations are not shown. Our probability calculations include SMBH locations where no extra images form. 

It can be seen from the map that there exists an elliptical `barrier' between the saddle point images (reddish points) and central images (greenish points), of radius $\sim$0.02". This barrier is the critical curve formed by the presence of SMBH. The closest images to this barrier are the brightest images generated from the model, while the further away the images become dimmer. The exact radius of this critical curve depends on the galaxy macro-model used. The right panel of Figure \ref{fig:RXJ1131all} shows the same image locations plotted onto the same galaxy macro-model's Fermat potential $\Psi_{\rm Fermat} = \frac{1}{2}r^2 - \Psi_{\rm gal}$. The asymmetric Fermat potential of this galaxy macro-model prefers that for the saddle and SMBH images to form, SMBH positions should be in the right of the field, as indicated by the density of saddle images in that region. In general, the shape of the Fermat potential determines which regions of the field are discriminated against having saddle and SMBH images, with more elliptical potentials strongly favoring SMBH locations for central images along the semi-major axis. The asymmetry of the mass distribution arising from quad-scale macro-modeling is important for central image properties, yet is often ignored in the literature.

For this model, the mean SMBH offset distance that produces extra central images is 0.282 $\pm$ 0.113 kpc. In general, the production of extra central images requires SMBH offsets greater than the mean $\Delta r$ from VODKA. This trend is important because the failure to create extra central images means the singular central image at the maximum of $\Psi_{\rm Fermat}$ will be strongly demagnified. 

Magnification statistics for this model are shown in Figure \ref{fig:RXJhistogram}. While in this case, $M_{\rm SMBH}$ is the same for all the models with the purpose of showcasing the observed magnification variability for a given $M_{\rm SMBH}$, our general procedure varies $M_{\rm SMBH}$ randomly for each model according to the distribution outlined in Section \ref{txt:imagegeneration}. As with our map in the left panel, SMBH locations where no central images form are excluded. In general, SMBH locations with no central images make up $\sim$50\% of all offset distances, meaning roughly half of all SMBH offsets will not yield observable images. The vertical dashed lines correspond to image magnitudes of $m_{\rm AB}=30$ and $34$ in the F218W HST filter. For this particular model we find that most ($\sim$62\%) of the central images are brighter than $m_{\rm AB}=30.00$. The corresponding saddle images have a much broader distribution and are generally much fainter than the central images. Combining each of these maps and magnification histograms for all the models in each QSO allows us to place statistical constraints on $r_c$ and $\Delta r$.
\begin{figure*}
    \centering
    \includegraphics[trim={5.5cm 0.35cm 5.42cm 0.37cm},clip,width=0.49\textwidth]{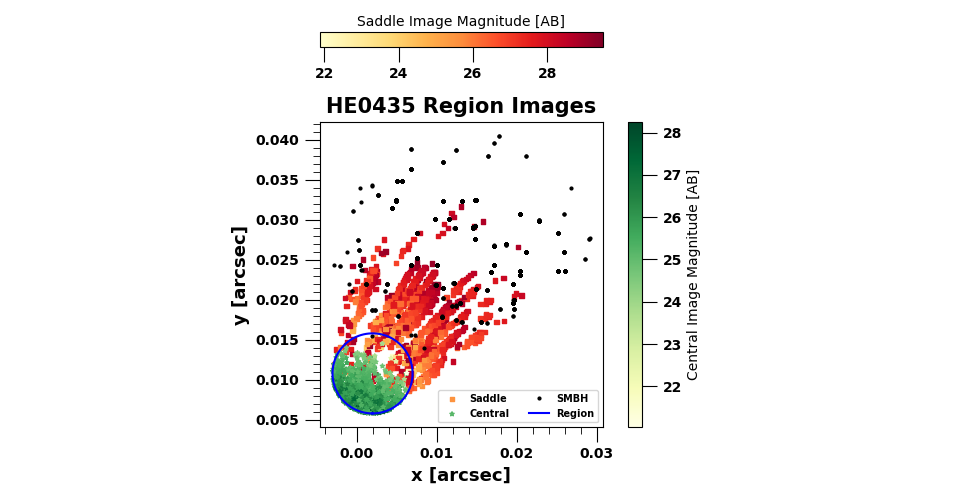}
    \includegraphics[trim={5.5cm 0.35cm 5.42cm 0.37cm},clip,width=0.49\textwidth]{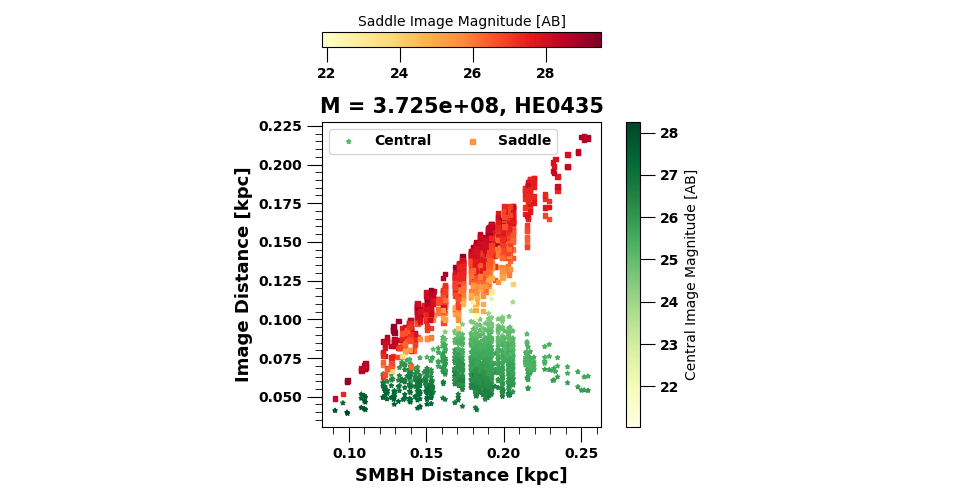}
\caption{{\it Left:} SMBH (all have $M_{\rm SMBH}\approx 3.73\times10^{8}M_\odot$) locations (black dots) corresponding to central images (green stars) and saddle images (orange squares) generated from a subset of 100 mass models for HE0435-1223 falling within the $r=0.0050"$ region (blue circle) centered at $(0.0020",0.0108")$ from the mass density peak. Orange squares depict simultaneously produced saddle images. The right and top colorbars indicate the magnitude of the central and saddle images, respectively. {\it Right:} Central (green stars) and saddle (orange squares) image distances from the mass density peak as a function of their corresponding SMBH offset distance for central images falling within the $r=0.0050"$ region of the top panel. All our maps have 1250 pix arcsec$^{-1}$. The vertical streak pattern indicates the range of image locations for each SMBH distance given by the 100 model subset. The right and top colorbars show the magnitude of the central and saddle images, respectively.
}
\label{fig:HE0435region}
\end{figure*}

Our statistical inference method (see Section~\ref{txt:statistics}) requires a likelihood function dependent on the magnitude and position of the predicted central image. As a visualization of a predicted detection, we define a small circular region within the lens plane such that any central images falling within the region are considered to be detected with magnitudes according to their respective magnification. The left panel of Figure~\ref{fig:HE0435region} depicts one example region for a subset of 100 HE0435-1223 galaxy macro-models, centered at $\vec r_o = (0.0020",0.0108")$ with a radius of 0.0050". For this lens, this radius corresponds to $\sim$0.03 kpc. The radius of the region can be thought of as the uncertainty $\sigma_r$ of a detection at $\vec r_o$. All central images found within the region and their corresponding saddle images and SMBH locations are shown. The streak pattern is due to the fact that saddle images form on the line connecting the center of the lensing potential and the SMBH, shown here as black points.  The SMBH offset distribution average is measured to be 0.187 kpc with a corresponding standard deviation of 0.017 kpc.

The right panel of Figure~\ref{fig:HE0435region} presents the information for the same circular region shown in the left panel, but plots the central and saddle image distances from the lensing potential peak against the corresponding SMBH distance. It is clear from the figure that the central and saddle images that form closer together, and hence closer to the critical curve, are brighter than those that form further away. Additionally, there is a general positive correlation between the SMBH position and its corresponding central and saddle image distances.

\subsection{Constraints on Lens Galaxy Parameters}\label{txt:constraints}

\begin{figure*}
\begin{multicols}{2}
    \includegraphics[width=\linewidth]{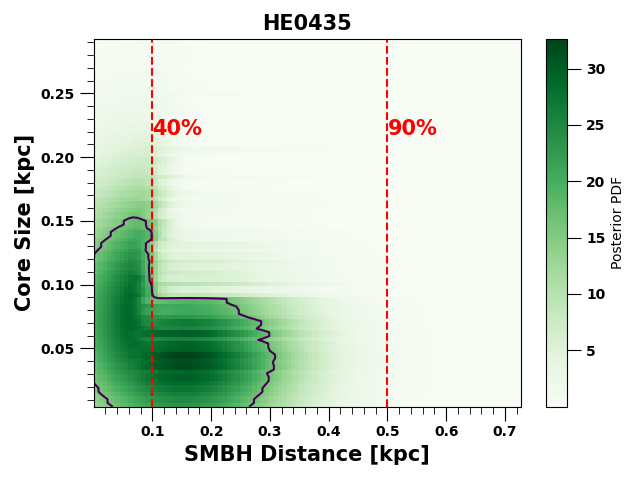}\par 
    \includegraphics[width=\linewidth]{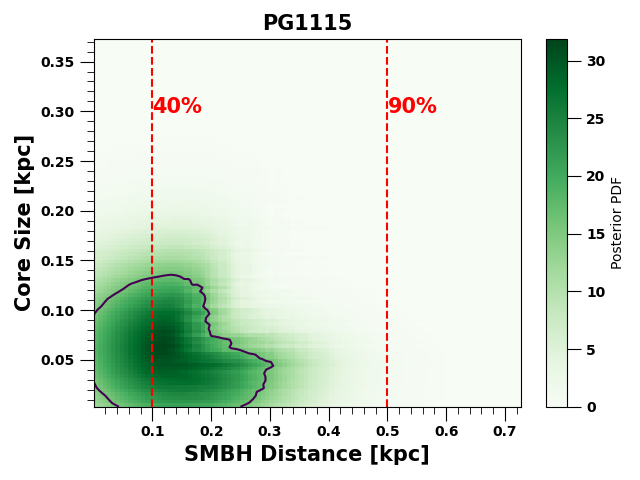}\par 
    \end{multicols}
\begin{multicols}{2}
    \includegraphics[width=\linewidth]{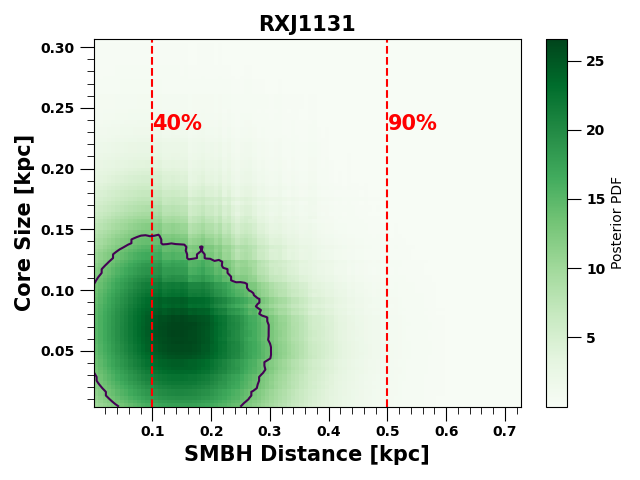}\par
    \includegraphics[width=\linewidth]{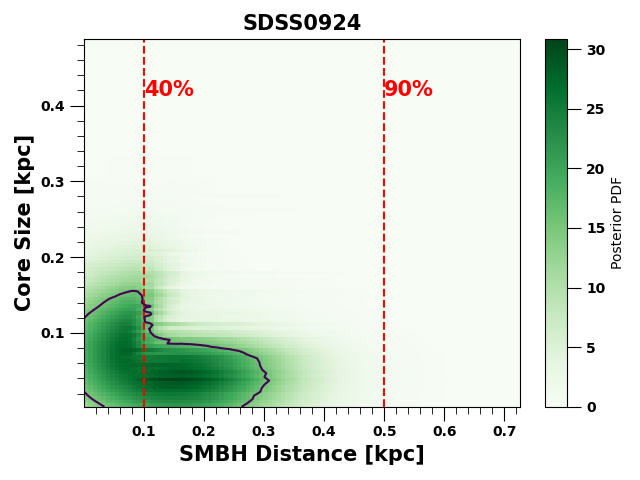}\par
\end{multicols}
\begin{multicols}{2}
    \includegraphics[width=\linewidth]{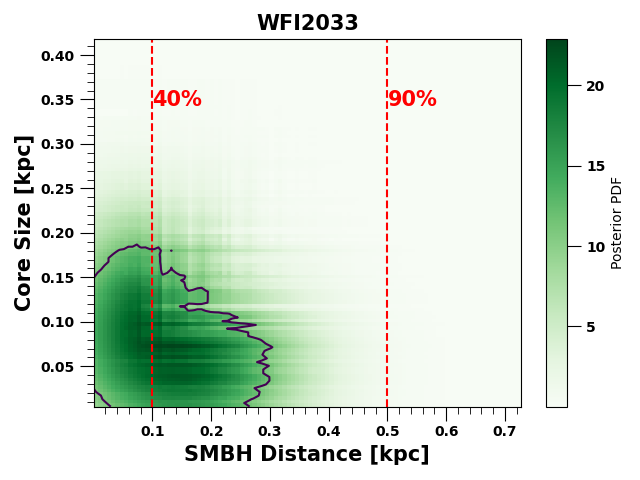}\par
    \includegraphics[width=\linewidth]{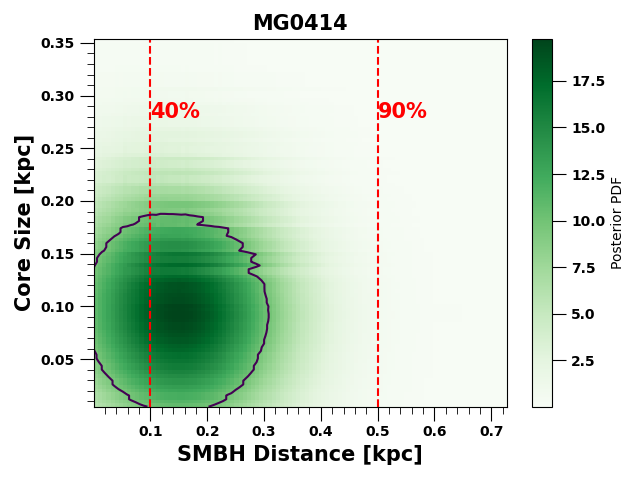}\par
\end{multicols}
\caption{Individual Joint Posterior PDFs $P(\Delta r, r_c | D)$ for the hypothetical case of a non-detection of the central image in all QSOs in our sample. Marginalizing over each axis gives individual posterior PDFs for core size $P(r_c | D)$ and SMBH offset distance $P(\Delta r | D)$. The vertical red dashed lines indicate the percentage of SMBH offset by $<$0.1 and $<$0.5 kpc from VODKA \citep{shen19}.}
\label{fig:indivnondetcontours}
\end{figure*}

\begin{figure*}
\begin{multicols}{2}
    \includegraphics[width=\linewidth]{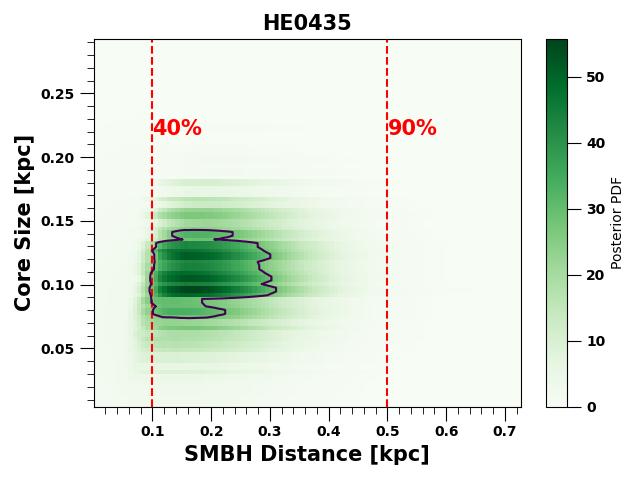}\par 
    \includegraphics[width=\linewidth]{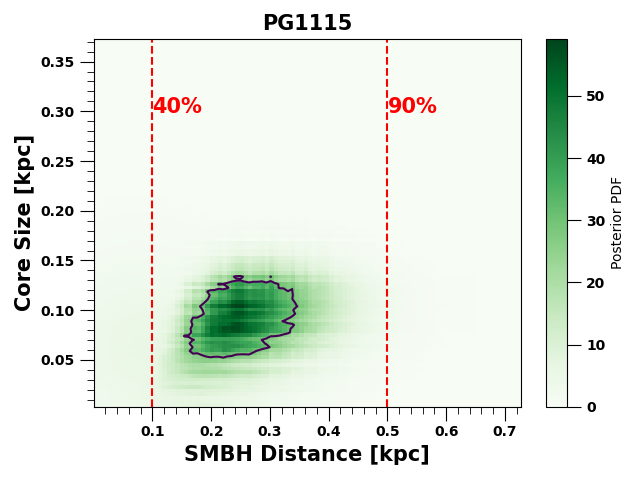}\par 
    \end{multicols}
\begin{multicols}{2}
    \includegraphics[width=\linewidth]{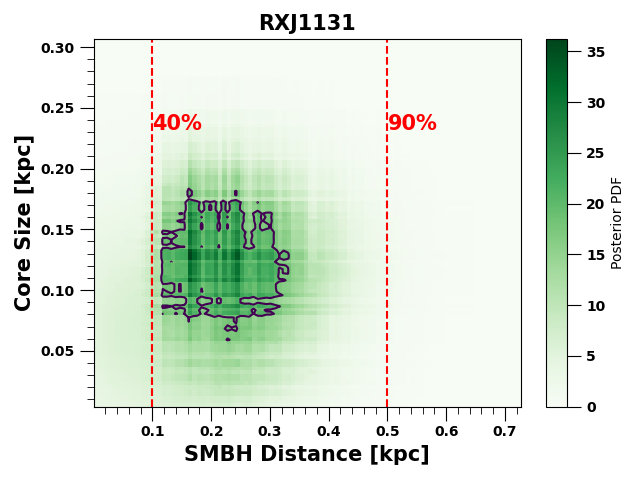}\par
    \includegraphics[width=\linewidth]{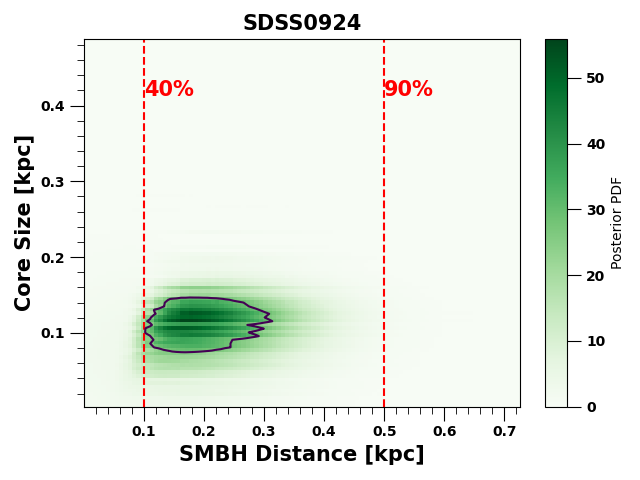}\par
\end{multicols}
\begin{multicols}{2}
    \includegraphics[width=\linewidth]{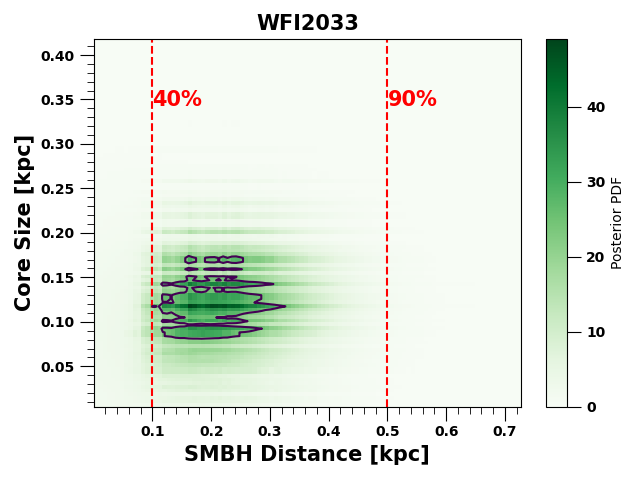}\par
    \includegraphics[width=\linewidth]{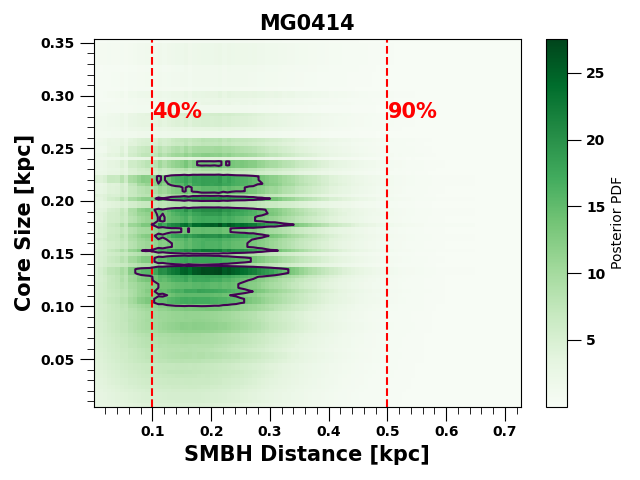}\par
\end{multicols}
\caption{Individual Joint Posterior PDFs $P(\Delta r, r_c | D)$ for the hypothetical case of a central image detection in all QSOs in our sample. The central image detection is assumed to be located at the center of each lens galaxy macro-model with a brightness of 9 magnitudes fainter than the brightest quad image in each lens. Marginalizing over each axis gives individual posterior PDFs for core size $P(r_c | D)$ and SMBH offset distance $P(\Delta r | D)$. The vertical red dashed lines indicate the percentage of SMBH offset by $<$0.1 and $<$0.5 kpc from VODKA \citep{shen19}.}
\label{fig:detcontours}
\end{figure*}

\begin{figure*}
\begin{multicols}{2}
    \includegraphics[width=\linewidth]{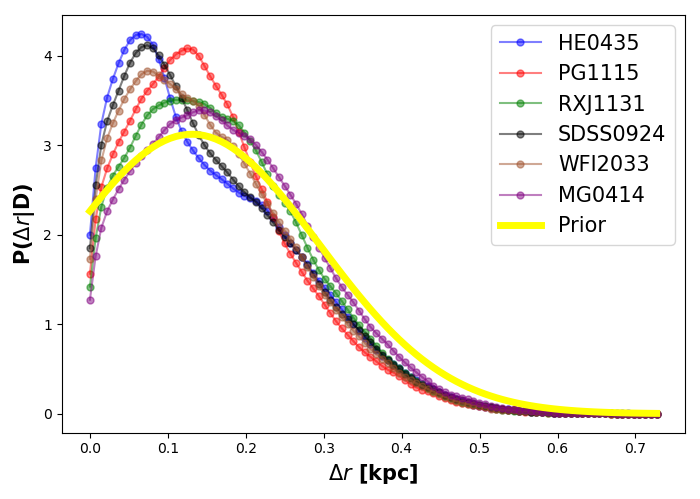}\par 
    \includegraphics[width=\linewidth]{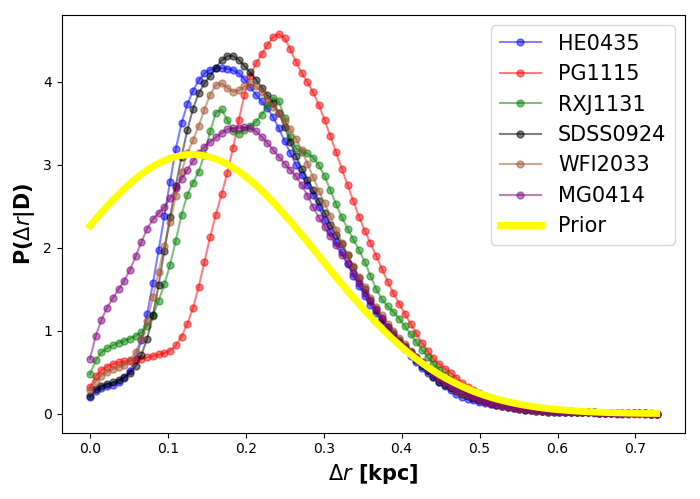}\par 
    \end{multicols}
\begin{multicols}{2}
    \includegraphics[width=\linewidth]{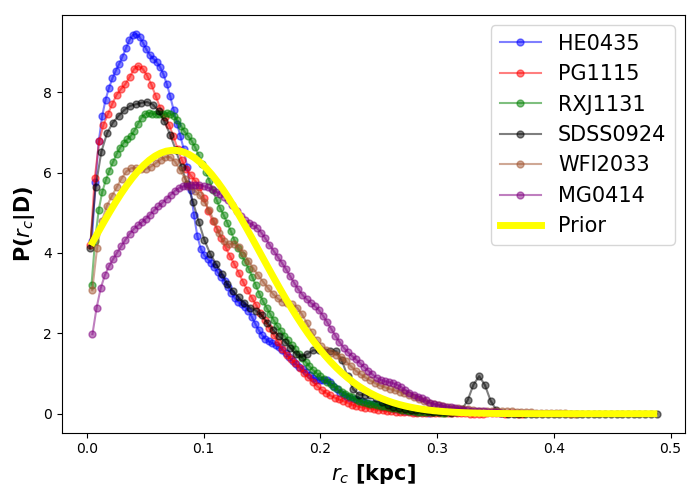}\par
    \includegraphics[width=\linewidth]{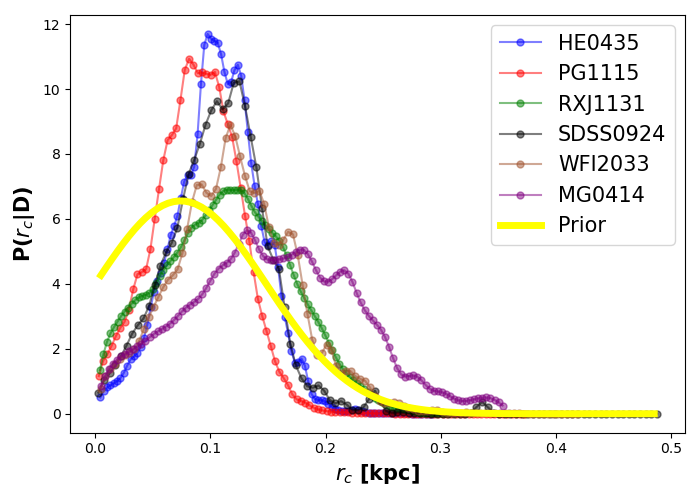}\par
\end{multicols}
\begin{multicols}{2}
    \includegraphics[width=\linewidth]{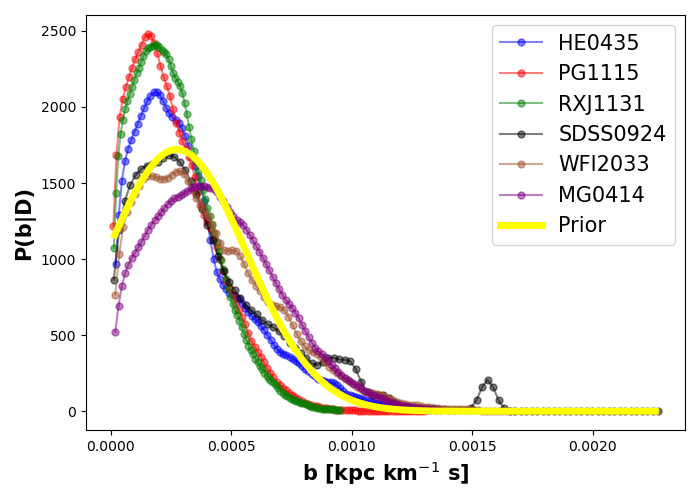}\par
    \includegraphics[width=\linewidth]{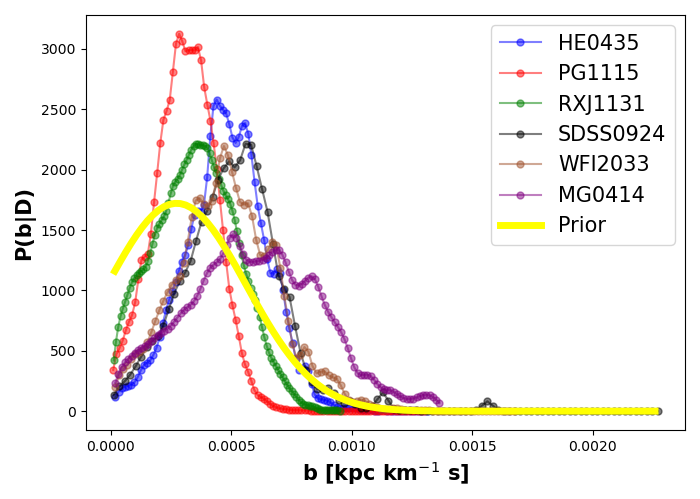}\par
\end{multicols}
\caption{Individual posterior PDFs for SMBH offset distance $\Delta r$ ({\it top row}), core size $r_c$ ({\it middle row})), and $r_c - \sigma$ proportionality constant $b$ ({\it bottom row}) for all QSOs in our sample. The case of a central image non-detection ({\it left column}) and detection ({\it right column}) yield independent constraints on lens galaxy parameters. The prior distributions for each parameter are shown in yellow. }
\label{fig:indivpost}
\end{figure*}

The previous section highlights how individual lens galaxy macro-models from our central image analysis can yield constraints on $\Delta r$ and $r_c$. This is largely the outcome of applying the analysis described in Section \ref{txt:imagegeneration}. The main goal of this work is to combine all models for each QSO in our sample to obtain general constraints on each lens galaxy's parameters. For this we follow the analysis procedure described in Section \ref{txt:statistics}. For each QSO, 1000 lens galaxy macro-models were generated. For each of these, and a randomly picked SMBH $\Delta r$ offset from the VODKA distribution, and randomly picked SMBH mass from the appropriate distribution,
we determined how many central images were produced, and each image's position and magnification. This allows us to create contour probability maps in $(\Delta r,r_c)$ space based on a predicted non-detection (see equation \ref{eq:nondetection}) or detection (see equation \ref{eq:detection}). We extend our statistical inference to 2 outcomes: (i) Non-detections and (ii) Detections. A summary of our constraints on lens galaxy parameter distributions for each quad is presented in Table \ref{tab:indivconstraints}. The posterior PDFs $P(\Delta r, r_c | D)$ for the cases of a non-detection and detection in each quad are shown in Figures \ref{fig:indivnondetcontours} and \ref{fig:detcontours}. Individual marginal posterior PDFs for each lens are shown in Figure \ref{fig:indivpost}.

\begin{table*}

	\caption{Constraints on Individual Lens Galaxy Parameters}
	\begin{tabular}{lllllllc} 
		\hline
		QSO (Det./Non-det.) & $\widehat{\Delta r}$ [kpc] & 95\% CI ($\Delta r$) [kpc] & $\widehat{r_c}$ [kpc] & 95\% CI ($r_c$) [kpc] &$\widehat{b}$ [10$^{-4}$ kpc km$^{-1}$ s] & 95\% CI ($b$) [10$^{-4}$ kpc km$^{-1}$ s] \\
		\hline
		HE0435-1223 (Non-det.) & 0.066 & $0.015 < \Delta r < 0.367$ & 0.042 & $0.010 < r_c < 0.179$ & 1.90 & $0.459 < b < 8.069$  \\
		HE0435-1223 (Det.) & 0.154 & $0.081 < \Delta r < 0.397$ & 0.095 & $0.042 < r_c < 0.165$ & 4.26 & $1.90 < b < 7.41$ &  \\
		PG1115+080 (Non-det.) & 0.118 & $0.015 < \Delta r < 0.353$ & 0.044 & $0.010 < r_c < 0.167$ & 1.54 & $0.365 < b < 5.832$ &  \\
		PG1115+080 (Det.) & 0.242 & $0.081 < \Delta r < 0.426$ & 0.081 & $0.025 < r_c < 0.145$ & 2.84 & $0.886 < b < 5.051$  \\
		RXJ1131-1231 (Non-det.) & 0.110 & $0.022 < \Delta r < 0.367$ & 0.056 & $0.013 < r_c < 0.181$ & 1.73 & $0.407 < b < 5.616$ &  \\
		RXJ1131-1231 (Det.) & 0.162 & $0.059 < \Delta r < 0.419$ & 0.126 & $0.022 < r_c < 0.206$ & 3.91 & $0.692 < b < 6.374$  \\
		SDSSJ0924+0219 (Non-det.) & 0.073 & $0.015 < \Delta r < 0.367$ & 0.037 & $0.008 < r_c < 0.213$ & 1.72 & $0.349 < b < 9.928$ &  \\
		SDSSJ0924+0219 (Det.) & 0.176 & $0.088 < \Delta r < 0.404$ & 0.125 & $0.032 < r_c < 0.174$ & 5.82 & $1.49 < b < 8.10$  \\
		WFI2033-4723 (Non-det.) & 0.073 & $0.015 < \Delta r < 0.360$ & 0.071 & $0.013 < r_c < 0.230$ & 2.86 & $0.518 < b < 9.041$ &  \\
		WFI2033-4723 (Det.) & 0.162 & $0.081 < \Delta r < 0.404$ & 0.117 & $0.034 < r_c < 0.209$ & 4.70 & $1.35 < b < 8.37$  \\
		MG0414+0534 (Non-det.) & 0.147 & $0.022 < \Delta r < 0.375$ & 0.089 & $0.019 < r_c < 0.234$ & 3.45 & $0.728 < b < 9.020$ &  \\
		MG0414+0534 (Det.) & 0.184 & $0.037 < \Delta r < 0.404$ & 0.132 & $0.033 < r_c < 0.273$ & 5.08 & $1.27 < b < 10.5$ &  \\
		RXJ0911+0551 (Non-det.) & 0.162 & $0.022 < \Delta r < 0.375$ & 0.485 & $0.084 < r_c < 0.868$ & 18.68 & $3.23 < b < 33.46$  \\		
		RXJ0911+0551 (Det.) & 0.184 & $0.022 < \Delta r < 0.389$ & 0.310 & $0.119 < r_c < 0.641$ & 11.96 & $4.57 < b < 24.72$  \\
		\hline
	\end{tabular}\\
\medskip{The columns list the QSO and constraint scenario (based on whether or not a central image is detected), modes for SMBH offset $\widehat{\Delta r}$, core size $\widehat{r_c}$, and $b$ parameter $\widehat{b}$, and 95\% credible intervals for SMBH offset, core size, and $b$ parameter. All modes and credible intervals presented are for the posterior PDF distributions for each parameter. RXJ0911+0551 is treated independently since its core size prior distribution is skewed to large core sizes due to the presence of large external shear.}
\label{tab:indivconstraints}
\end{table*}

\begin{table*}
	\caption{Constraints on Lens Galaxy Parameters}
	\begin{tabular}{lllllllc} 
		\hline
		Constraint Scenario & $\widehat{\Delta r}$ [kpc] & 95\% CI ($\Delta r$) [kpc] & $\widehat{r_c}$ [kpc] & 95\% CI ($r_c$) [kpc] &$\widehat{b}$ [10$^{-4}$ kpc km$^{-1}$ s] & 95\% CI ($b$) [10$^{-4}$ kpc km$^{-1}$ s] \\
		\hline
		All Detections & 0.242 & $0.154 < \Delta r < 0.286$ & 0.107 & $0.083 < r_c < 0.136$ & 6.28 & $4.47 < b < 7.87$  \\
		All Non-detections & 0.132 & $0.037 < \Delta r < 0.228$ & 0.049 & $0.019 < r_c < 0.107$ & 3.11 & $0.849 < b < 5.830$ &  \\
		Some Det./Non-det. & 0.191 & $0.110 < \Delta r < 0.264$ & 0.068 & $0.033 < r_c < 0.102$ & 4.47 & $2.21 < b < 6.96$  \\
		\hline
	\end{tabular}\\
\medskip{The columns list the constraint scenario (based on whether or not a central image is detected), modes for SMBH offset $\widehat{\Delta r}$, core size $\widehat{r_c}$, and $b$ parameter $\widehat{b}$, and 95\% credible intervals for SMBH offset, core size, and $b$ parameter. All modes and 95\% credible intervals presented are for the marginalized posterior PDF distributions for each parameter assuming the entire sample is governed by a single constraint value for $\Delta r$, $r_c$, and $b$}. The scenario for Some Det./Non-det. assumes a central image detection in PG1115+080 and WFI2033-4723 and non-detections in the rest.

\label{tab:constraints}
\end{table*}

Let us first consider the most likely case of central image non-detections. Our inference method in this case simply checks the probability of an image being brighter than the $f_{\rm crit}$ threshold (see equation \ref{eq:nondetection}). The posterior PDFs (see equation \ref{eq:post1}) for each lens in this case is shown in Figure \ref{fig:indivnondetcontours}. The case of non-detections prefers smaller $\Delta r$ values than the case of detections. In general, we can say that non-detections of central images imply that the central SMBH is very likely to be well centered on the galaxy's lensing potential. As with $\Delta r$, smaller values of $r_c$ and $b$ are strongly favored for the case of non-detections. Qualitatively, this is a similar conclusion to those of previous works \citep[e.g.][]{quinn16}.

Next, we consider the unlikely, but more interesting, hypothetical outcome that the central image was detected in the 6 lens systems in our sample. The resulting posterior PDFs are shown in Figure \ref{fig:detcontours}. We can see that in the event of a detection at the chosen $\vec r_o$ with $f_o$ in each lens system, $\Delta r$ is most likely to be $>$0.1 kpc. The case of detections favors larger $\Delta r$. Similarly for $r_c$, the case of detections constrains to larger core sizes and $b$ values.

\begin{figure}
    \centering
    \includegraphics[trim={ 0.15cm 0.35cm 0cm 0cm},clip,width=0.49\textwidth]{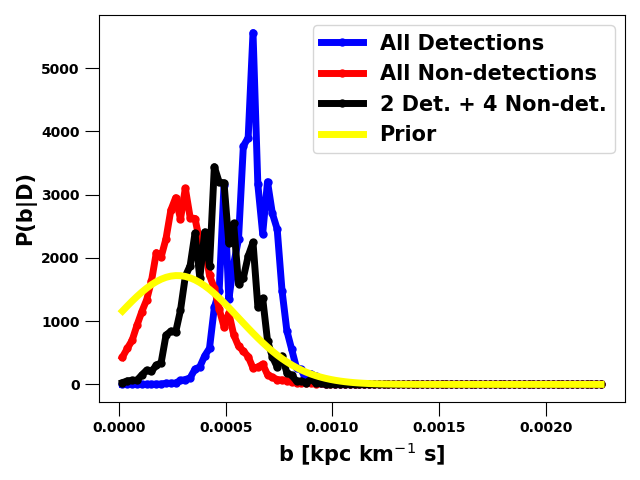}    
\caption{ Combined posterior PDFs for $r_c - \sigma$ proportionality constant $b$. The constraints from a hypothetical central image detection in each QSO are shown in blue. The constraints from non-detections of the central image in each QSO are shown in red. Constraints from a central image detection in PG1115+080 and WFI2033-4723 but non-detections in all others are shown in black. The prior distribution is shown in yellow. In general, non-detections of the central image favor smaller $b$ values.
}
\label{fig:allpdfs_b}
\end{figure}

Lastly, since the $b$ parameter is assumed to be a global parameter for all lenses, we can multiply the 2D posteriors together and marginalize according to equation \ref{eq:post4} to get a general constraint on $b$. We do this for three scenarios: (i) All Non-detections in the sample, (ii) All detections in the sample, and (iii) A combination of Non-detections and detections. For this third case, we assumed that the lenses with the top two brightest observed quad images, PG1115+080 and WFI2033-4723, had hypothetical central image detections. With this assumption, we multiplied together individual non-detection posteriors for the rest of the lens systems in our sample with the detection posteriors for the two chosen lens systems that we promoted to detections. Each scenario yields independent constraints on $b$. The summary of these constraints is detailed in Table \ref{tab:constraints} and Figure \ref{fig:allpdfs_b}. In general, central image detections favor steeper $r_c - \sigma$ ($b = 6.28^{+1.59}_{-1.81} \times 10^{-4}$ kpc km$^{-1}$ s), while the inverse is true for non-detections ($b = 3.11^{+2.72}_{-2.26} \times 10^{-4}$ kpc km$^{-1}$ s). The combination case yields $b = 4.47^{+2.49}_{-2.26} \pm 1.40 \times 10^{-4}$ kpc km$^{-1}$ s, between the detection and non-detection case constraint. In fact, in each of the three scenarios, the resulting distributions for $b$ have credible intervals smaller than that of the prior, indicating that the search for central images with this analysis can yield stronger constraints regardless of outcome.

\subsection{Rough Constraints for $\Delta r$ and $r_c$} \label{txt:rough}

\begin{figure}
    \centering
    \includegraphics[trim={ 0.15cm 0cm 0cm 0cm},clip,width=0.49\textwidth]{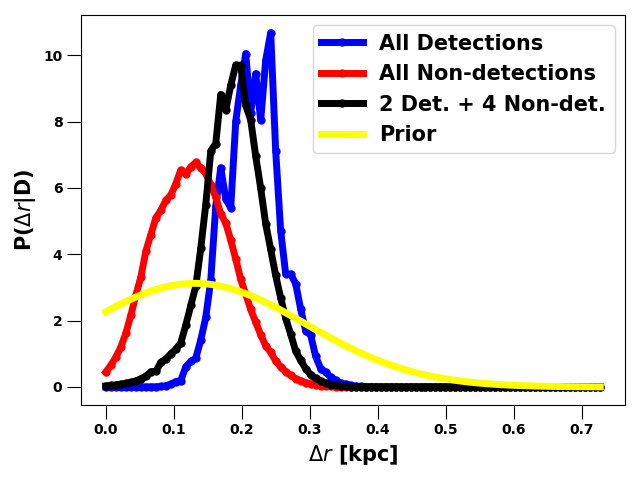}
    \includegraphics[trim={ 0.15cm 0cm 0cm 0cm},clip,width=0.49\textwidth]{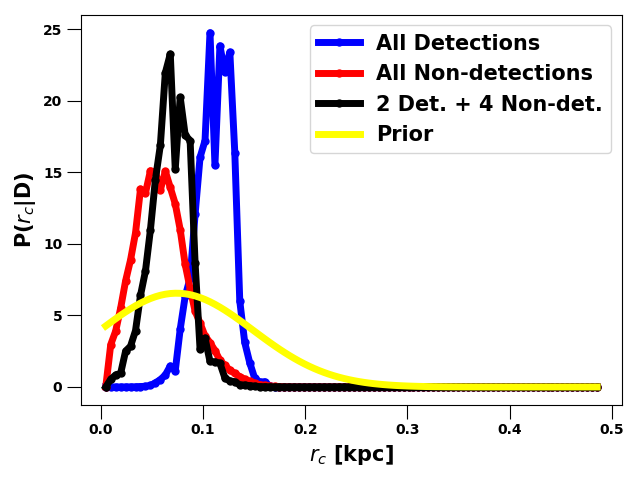}
\caption{ Combined posterior PDFs for SMBH offset distance $\Delta r$ ({\it top panel}) and core size $r_c$ ({\it bottom panel}) assuming these parameters are the same for all lens galaxies}. The constraints from a hypothetical central image detection in each QSO are shown in blue. The constraints from non-detections of the central image in each QSO are shown in red. Constraints from a central image detection in PG1115+080 and WFI2033-4723 but non-detections in all others are shown in black. The prior distributions for each parameter are shown in yellow. In general, non-detections of the central image favor smaller SMBH offset distance and core sizes. 

\label{fig:allpdfs}
\end{figure}

\begin{figure}
    \centering
    \includegraphics[trim={ 0.15cm 0.35cm 0cm 0cm},clip,width=0.49\textwidth]{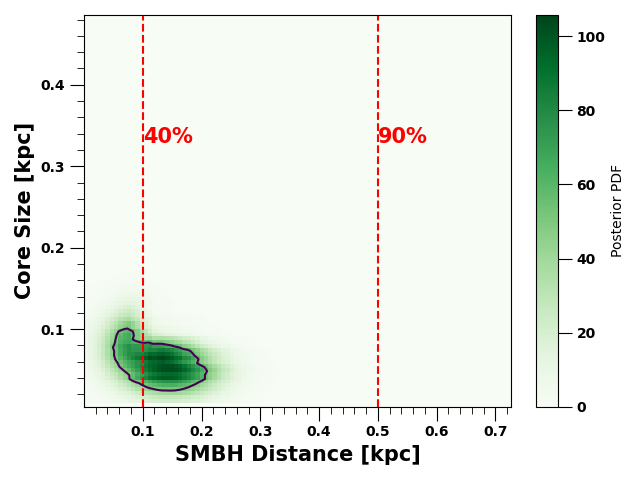}
    \includegraphics[trim={ 0.15cm 0.35cm 0cm 0cm},clip,width=0.49\textwidth]{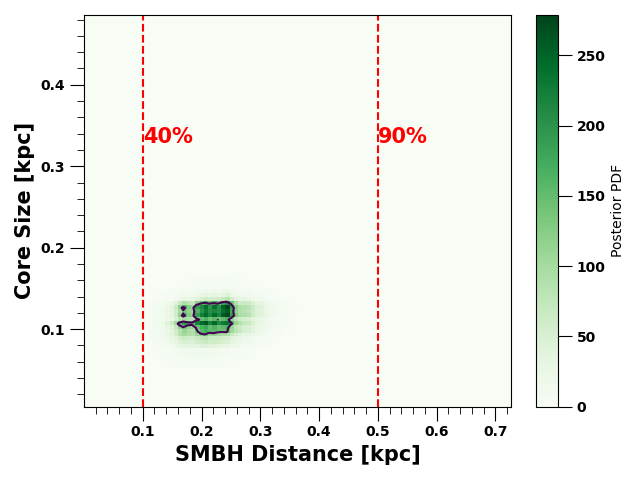}
    \includegraphics[trim={ 0.15cm 0.35cm 0cm 0cm},clip,width=0.49\textwidth]{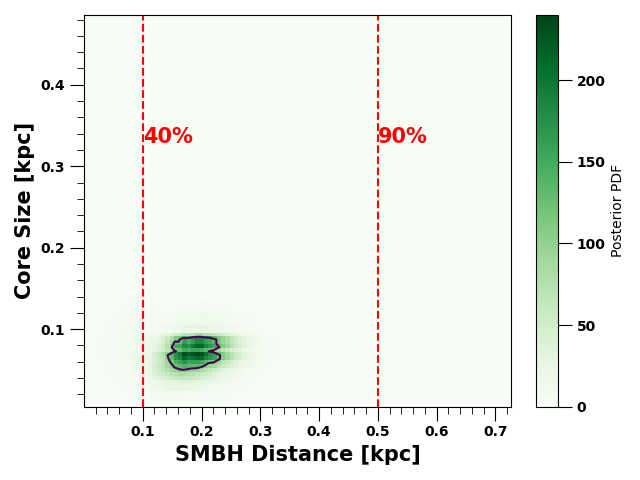}
\caption{ Posterior PDFs $P(\Delta r, r_c | D)$ for the hypothetical case of central image non-detections in all QSOs in our sample ({\it top panel}), central image detections in all QSOs in our sample ({\it middle panel}), and a central image detection in PG1115+080 and WFI2033-4723, and non-detections in all other QSOs in our sample ({\it bottom panel}). For detections, the central image detection is assumed to be located at the center of each lens galaxy macro-model with a brightness of 9 magnitudes fainter than the brightest quad image in each lens. Marginalizing over each axis gives individual posterior PDFs for core size $P(r_c | D)$ and SMBH offset distance $P(\Delta r | D)$, assuming $r_c$ and $\Delta r$ are the same for all lens galaxies. The vertical red dashed lines indicate the percentage of SMBH offset by $<$0.1 and $<$0.5 kpc from VODKA \citep{shen19}.
}
\label{fig:totcontours}
\end{figure}

Our results so far have shown that the non-detection or detection of a central image can constrain the lens galaxy SMBH offset and core radius. To place a general constraint on these parameters, we assumed a proportionality with $r_c$ and $\sigma$ and marginalized over $\Delta r$ to obtain the posterior for this proportionality constant $b$. We are able to do this since $b$ is a parameter that is assumed to be the same for all lens galaxies. In this section, we similarly assume that all lens galaxies share the same SMBH offset and core radius in order to directly constrain $\Delta r$ and $r_c$. Since this is an unrealistic assumption, we treat these results as a "back of envelope" estimation rather than a strict constraint. In fact, since the estimated $\Delta r$ and $r_c$ for individual lens galaxies are approximately similar to one another (see Table \ref{tab:indivconstraints}), this exercise serves as a useful estimate of the rough scale of SMBH offsets and core radii that one can expect in future analyses.

With this assumption established, we follow equations \ref{eq:post3} and \ref{eq:combinepostrc} to derive posterior PDFs for $\Delta r$ and $r_c$, respectively. We do this for the same 3 scenarios as for the constraint on $b$ (All Non-detections, All Detections, and a Combination of Detections and Non-detections. As before, we assume a detection in PG1115+080 and WFI2033-4723 but non-detections in all others for the third case.). A summary of all the posterior PDF constraints for $\Delta r$ and $r_c$ is shown in Figure \ref{fig:allpdfs} and Table \ref{tab:constraints}. Similarly, the total multiplied together 2D posteriors for each case are shown in Figure \ref{fig:totcontours}. 

For all non-detections in our sample, the mode of the posterior $P(\Delta r | D)$  is 0.132 kpc with a 95\% credible interval of $0.037 < \Delta r < 0.228$ kpc. This distribution is skewed to smaller $\Delta r$, consistent with the trend that non-detections of central images imply that the central SMBH is very likely to be well centered on the galaxy's lensing potential. With core size, the $r_c$ posterior mode is 0.049 kpc. For all detections, the $\Delta r$ posterior mode is 0.242 kpc with a 95\% credible interval of $0.154 < \Delta r < 0.286$ kpc. This is a stronger constraint than that of the case of all non-detections. the $r_c$ posterior mode for all detections is 0.107 kpc. In the combination case, the constraints on $\Delta r$, $r_c$, and $b$ lie in between those found in the cases of non-detections and detections. This intuitively makes sense as overall, central image detections favor larger SMBH offsets and core sizes and steeper $r_c - \sigma$, while the inverse is true for non-detections. In addition, the posterior means for each parameter are similar to the prior means, but with smaller credible intervals.

\section{A New Detection Technique}\label{txt:detection}

The results of our analyses can be applied to future observations of the elusive central image in hopes of constraining various galaxy properties. Here we outline a novel observing strategy that can help improve the chances of central image detection. 

In addition to being significantly demagnified, the central image is often superimposed by the lens galaxy in optical wavelengths. Therefore, to increase the likelihood of detecting the central image, it is important to increase the brightness contrast between it and the lens galaxy,  
that is, to maximize the central image flux and minimize the lens galaxy flux.  To achieve that, the observed wavelength of the galaxy should be shorter than the rest frame wavelength of the Balmer break ($3646\,$\AA). The lens galaxies will be largely invisible in these wavelengths, because they are predominantly old, massive ellipticals, and thus have spectral energy densities that peak in red wavelengths and drop off toward the blue and ultraviolet.
Furthermore, QSO sources have a blue power-law continuum, with high flux densities in bluer wavelengths. Additionally, to avoid the wavelength range absorbed by intervening intergalactic medium (Ly-$\alpha$ forest), the observed wavelength of the QSO should be greater than the rest frame wavelength of $\sim1200\,$\AA, which means the source QSOs should have redshifts above $\sim 1.5$ \citep[][]{francis91,hewett10}. In practice, this only becomes important for sources at redshifts greater than $\sim$2.
This implies that the ideal scenario for which this technique can be utilized is the highest possible redshift for the lens galaxy for any given source redshift.

In order to accomplish this, we suggest observing in the bluest wavelengths achievable with the UV filters on the Hubble Space Telescope (HST). In blue filters, the contrast of the lens galaxy and central image can be increased to the point where the central image can become detectable. 

In addition to brightness considerations, the positioning of the central image is better constrained for compact lens sources, which is another reason why QSOs are ideal sources.

\begin{figure*}
    \centering
    \includegraphics[width=0.49\textwidth,height=6cm]{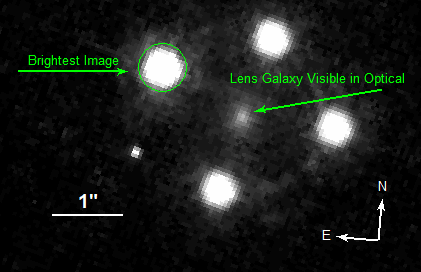}
    \includegraphics[width=0.49\textwidth,height=6cm]{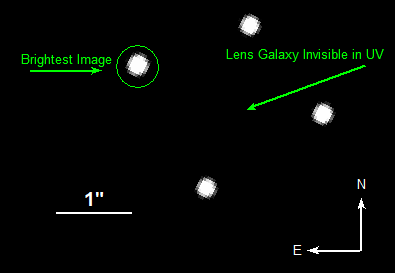}
\caption{ HE0435-1223 in HST optical filter F555W ({\it left panel}) and UV filter F275W ({\it right panel}). Lens and source redshifts are $z_l=0.455$ and $z_s=1.693$ for this system. In F555W, the lensing galaxy is visible in the center of the image since its rest frame emission in this filter lies just longward of Balmer break. In F275W, the lensing galaxy is not visible, while the QSO images are bright, showing how UV searches for the central image can be viable. $1"=5.8\,$kpc at the redshift of the lens.
}
\label{fig:limits}
\end{figure*}

As a preliminary illustration of this technique, we consider the HST UV filter F275W and Optical filter F555W, which have effective wavelengths\footnote{Effective wavelength values taken from the SVO Filter Profile Service: \url{http://svo2.cab.inta-csic.es/theory/fps/}} of $2713.86\,$\AA~ and $5326.96\,$\AA, respectively. Observing the gravitationally lensed QSO quad HE0435-1223 (properties presented in Table \ref{tab:qso_sample}) in F555W corresponds to rest frame QSO and lens emission at $\sim\!1980\,$\AA~ and $\sim\!3664\,$\AA~, respectively. In this filter, the lens emission is longward of the Balmer break limit, so the lens galaxy would obscure the central image. However, observing in F275W would correspond to rest frame emission at $\sim\!1008\,$\AA~ and $\sim\!1866\,$\AA~ for the QSO and lens, respectively. This wavelength corresponds to a regime where the QSO continuum flux is large. Similarly, the lens rest frame is in the regime that should have minimal flux from the lens galaxy. Therefore, our central image detection technique can be potentially viable for this system, as illustrated in Figure~\ref{fig:limits}.

\section{Discussion and Conclusions}\label{txt:conclusions}

Using a sample of 7 quad lenses, we modelled the central image and placed constraints on lens galaxy parameters for various scenarios of central image detection. We estimated 3 lens galaxy parameters: $b$ (the $r_c - \sigma$ proportionality constant), SMBH offset $\Delta r$, and galaxy core radius $r_c$. Constraints on these parameters for individual lenses are presented in Table \ref{tab:indivconstraints}.

Here we concentrate on the combined constraints for all lenses, assuming roughly that $\Delta r$ and $r_c$ are the same for all galaxies, while $b$ is defined to be this way. Our main results and constraints are presented in Table \ref{tab:constraints}, and are as follows:
\begin{itemize}
    \item All the cases we considered are hypothetical. However, the one that comes closest to the current observational constraints is the case with no detections of the central image in any of the lenses (red curves in Figure~\ref{fig:allpdfs}). For these, the SMBH offset from the center of the galaxy host is 132$^{+96}_{-95}$~pc, the galaxy core radius is 49$^{+58}_{-30}$~pc, and the constant of proportionality $b$ (eq.~\ref{eq:rsigma}), relating the core radius and the galaxy line-of-sight velocity dispersion is $3.11^{+2.72}_{-2.26} \times 10^{-4}$ kpc km$^{-1}$ s. Each of the constraint distributions yield much tighter constraints for each parameter than their respective priors, and favor smaller values.
    \item The more exciting, but unlikely, case of a central image detection in each lens also yields tight constraints on lens galaxy parameters (blue curves in Figure \ref{fig:allpdfs}). For these, the SMBH offset from the galaxy host center is 242$^{+44}_{-88}$ pc, the galaxy core radius is 107$^{+29}_{-24}$ pc, and the average constant of proportionality $b$, relating the core radius and the galaxy line-of-sight velocity dispersion is $6.28^{+1.59}_{-1.81} \times 10^{-4} $ kpc~km$^{-1} $s. A central image detection will therefore imply that the galaxy core sizes and SMBH offsets are larger than the average of their respective priors.
    \item The case of a combination of detections and non-detections in our sample yields independent constraints on lens galaxy parameters. With our assumed hypothetical detection in PG1115+080 and WFI2033-4723, justified by their bright quad images, the average SMBH offset, galaxy core radius, and constant of proportionality relating the core radius and the galaxy line-of-sight velocity dispersion are 191$^{+73}_{-81}$ pc, 68$^{+34}_{-35}$ pc, and $4.47^{+2.49}_{-2.26} \times 10^{-4} $ kpc~km$^{-1} $s, respectively. In general, even just a single detection of the central image will strongly influence the constraints to larger SMBH offset, core radius, and $b$.
    \item Regardless of whether or not the central image is detected in a lens or not, tighter constraints can be placed on SMBH offset, galaxy core radius, and the $r_c - \sigma$ proportionality constant. A notable exception to this result is exemplified in RXJ0911+0551 (see Section \ref{txt:rxj0911}). Due to the presence of a nearby galaxy cluster which leads to a wide radial range of its image distribution, and requires a large external shear, RXJ0911+0551 does not constrain any lens galaxy parameter regardless of detection. From this, it is likely that lenses that require large external shear due to the presence of nearby external mass will not be useful in constraining lens galaxy parameters in future studies.
    \item Our analysis demonstrates that the quad-scale lens macro-model is important for deriving properties of the central images. This is especially true in the case of very asymmetric systems, like RXJ0911+0551.
    \item While our results are not restricted to any particular observational filter, we recommend a novel observing strategy utilizing UV wavelength filters (see Section \ref{txt:detection}). Given that the lens galaxy peaks in red wavelengths and the QSO source peaks in blue wavelengths, UV filters have a potential use in searches for central images since they will obscure the lens galaxy and possibly allow for easier detection of the elusive demagnified blue central image. 
\end{itemize}

In this paper we considered how the fluxes of central QSO images are affected by the presence of SMBH, which can be offset from the center of its galaxy host. Because the central galaxy region is dominated by stars, the central image flux can also be affected by stellar microlensing \citep{dob07}, which would broaden the predicted flux distribution by $\sim$1 mag. Future studies should continue to account for this in central image modelling.

While our rough calculations for a single value of $\Delta r$ and $r_c$ for all galaxies (see Section \ref{txt:rough}) are useful in understanding the general scale of their underlying distribution, a more insightful result would be direct constraints on the distributions of $\Delta r$ and $r_c$. This would require a more complicated theoretical and statistical model on top of our outlined framework, and would be the subject of future study. 

Future observations with the Vera C. Rubin Observatory, Gaia, and Pan-Starrs will discover many new lensed systems \citep{marshall10,lemon19,canameras20} allowing the extension of our analysis to a larger sample, giving tighter constraints, and increasing the probability of detecting a central image. Similarly, combining our constraints with recent results from gravitational lensing of source AGN at $z > 1$ \citep{millon22,spingola22} can extend our constraints to the inner regions of higher redshift galaxies. Furthermore, our results can be compared with simulations \citep{tremmel18,volonteri20,katz20}. 

In this work we have shown that searches for the central image in quads can yield tight constraints on lens galaxy parameters regardless of detection or non-detection. Therefore, we recommend commencing new observation programs for the central image to formalize new constraints for SMBH offset, galaxy core radius, and $r_c - \sigma$ proportionality constant.

\section*{Acknowledgements}

The authors would like to thank Lindsey Gordon, Galin Jones, John Hamilton Miller Jr., and Sarah Taft for useful suggestions and discussions. 


\section*{Data Availability}

Data generated from this article will be shared upon reasonable request to the corresponding author.





\appendix

\section{The Special Case of RXJ0911+0551}\label{txt:rxj0911}

\begin{figure*}
    \centering
    \includegraphics[trim={4.0cm 0.35cm 4.5cm 0.37cm},clip,width=0.49\textwidth]{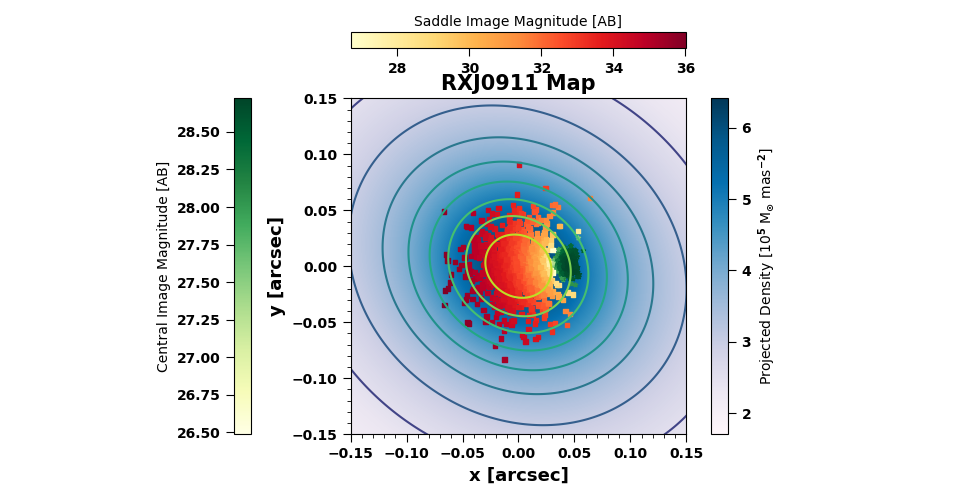}
    \includegraphics[trim={4.0cm 0.35cm 4.5cm 0.37cm},clip,width=0.49\textwidth]{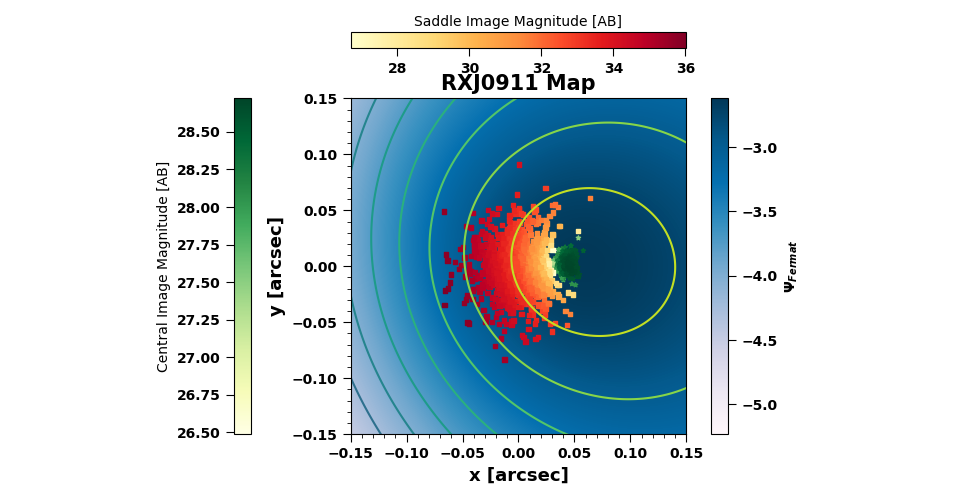}
\caption{ Central image modelling for RXJ0911+0551 quad lens. We used 1576 SMBHs (not shown) randomly offset from the galaxy host center according to the distribution in \citet{shen19}. In this figure all SMBH have $M_{\rm SMBH}\approx 9.03\times10^{8}M_\odot$. (Other maps for this system use different $M_{\rm SMBH}$ randomly chosen from its range shown in Table \ref{tab:qso_sample}).  Offset SMBHs that produce central images (green stars) and saddle images (inverted parity; orange squares) are plotted on the mass density {\it (left panel)} and Fermat potential $\Psi_{\rm Fermat}$ ({\it right panel}) of one model of RXJ0911+0551 ($\Psi_{\rm SMBH}$ is not included so as to avoid cluttering the plot). Cases where only one central image is produced at the location of the offset SMBH are excluded from this plot, but included in subsequent calculations. The left and top colorbars indicate the magnitude of the central and saddle images, respectively. The maps have scales of 0.0059 kpc pix$^{-1}$. All our maps have 1250 pix arcsec$^{-1}$.
}
\label{fig:RXJ0911}
\end{figure*}

\begin{figure*}
    \centering
    \includegraphics[trim={ 0.15cm 0.35cm 0cm 0cm},clip,width=0.49\textwidth]{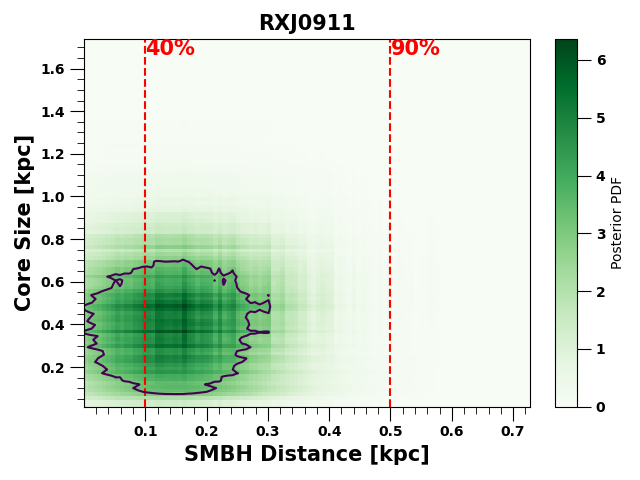}
    \includegraphics[trim={ 0.15cm 0.35cm 0cm 0cm},clip,width=0.49\textwidth]{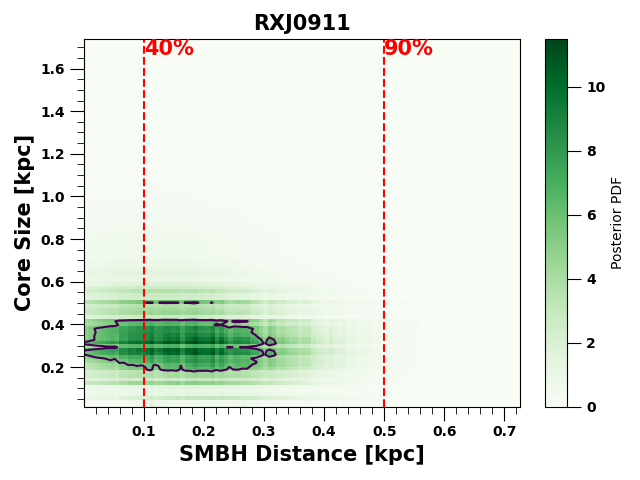}
\caption{ Posterior PDF, $P(\Delta r, r_c | D)$ for the hypothetical case of a central image non-detection ({\it left panel}) and detection ({\it right panel}) in RXJ0911+0551. Note that the vertical axis extends to larger core radii than in Figures~\ref{fig:indivnondetcontours} and \ref{fig:detcontours}. The detected central image is assumed to be located at the center of each lens galaxy macro-model, with a brightness of 9 magnitudes fainter than the brightest quad image in each lens. Marginalizing over each axis gives individual posterior PDFs for core size $P(r_c | D)$ and SMBH offset distance $P(\Delta r | D)$. The vertical red dashed lines indicate the percentage of SMBH offset by $<$0.1 and $<$0.5 kpc from VODKA \citep{shen19}. 
}
\label{fig:combocontoursRXJ}
\end{figure*}

\begin{figure}
    \centering
    \includegraphics[trim={ 0.15cm 0cm 0cm 0cm},clip,width=0.49\textwidth]{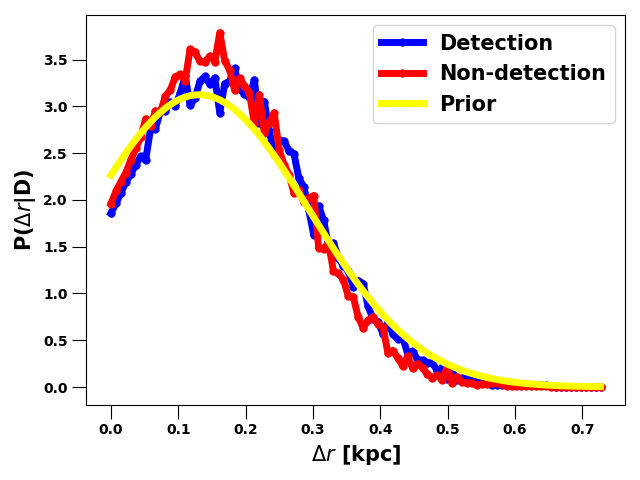}
    \includegraphics[trim={ 0.15cm 0cm 0cm 0cm},clip,width=0.49\textwidth]{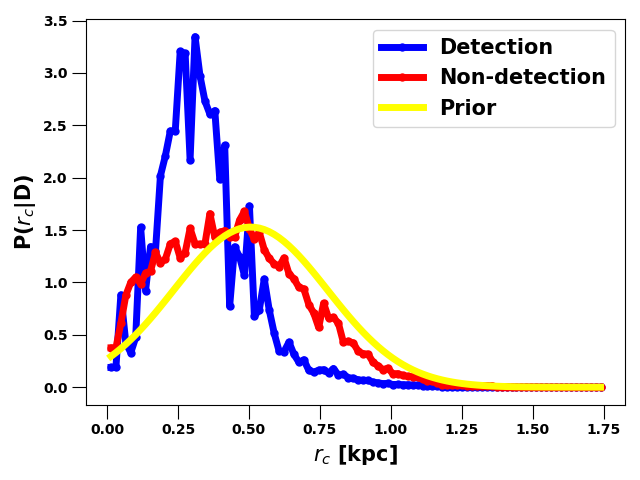}
    \includegraphics[trim={ 0.15cm 0.35cm 0cm 0cm},clip,width=0.49\textwidth]{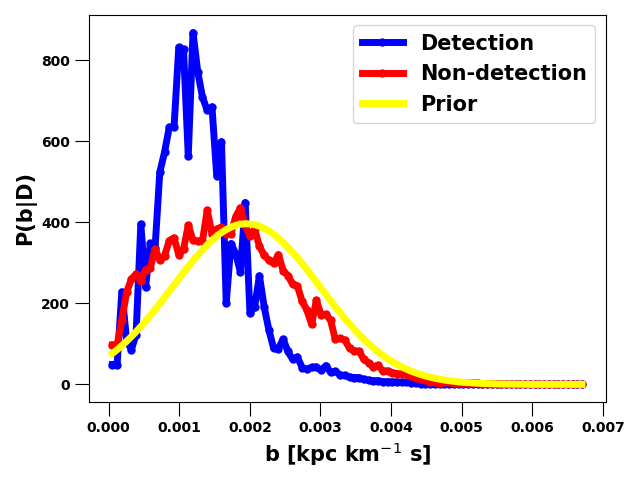}    
\caption{ RXJ0911+0551 posterior PDFs for SMBH offset distance $\Delta r$ ({\it top panel}), core size $r_c$ ({\it middle panel}), and $r_c - \sigma$ proportionality constant $b$ ({\it bottom panel}). The constraints from a hypothetical central image detection (non-detection) in this lens are shown in blue (red). The prior distributions for each parameter are shown in yellow. 
}
\label{fig:RXJ0911pdfs}
\end{figure}

RXJ0911+0551 is influenced by a large external shear from a nearby cluster. Applying the same initial range of quad-scale galaxy parameters as for the other 6 quads, but a somewhat larger external shear (see Section~\ref{txt:modelgeneration}) results in the core size distribution $P(r_c)$ to be skewed to much larger values than the rest of the sample. To compensate, we analyze it independently to get separate constraints on $\Delta r$, $r_c$, and $b$. The influence of the external shear can be seen in the lens mass-model maps in Figure \ref{fig:RXJ0911}. Compared with the lens maps for RXJ1131-1231 (see Figure \ref{fig:RXJ1131all}), which is the system most similar to RXJ0911+0551 of the 6 we use, we can easily see that the region of the central images (green stars) is displaced from the center of the galaxy density map further in RXJ0911+0551 than in RXJ1131-1231. Since the VODKA distribution is with respect to the galaxy center, the large external shear and offset position of the source displace the maximum of $\Psi_{\rm Fermat}$ further from the center, leading to small offset SMBH being able to produce extra images (this did not occur in the other lenses with centrally symmetric $\Psi_{\rm Fermat}$). 

The constraints for this system are listed in Table \ref{tab:indivconstraints}. The posterior PDFs $P(\Delta r, r_c | D)$ for the cases of a non-detection and detection are shown in Figure \ref{fig:combocontoursRXJ}. For a central image non-detection, the posterior PDF appears similar with the detection posterior, except with smaller core sizes allowed. The non-detection posterior PDF for $\Delta r$ has a mode of 0.162 kpc and 95\% credible interval of $0.022 < \Delta r < 0.375$ kpc. 

It is important to note that with RXJ0911+0551 there is significant overlap in constraints from detections and non-detections for $\Delta r$, something that is absent in the main sample (see Figure \ref{fig:indivpost}). The two regions overlap because the parameter that differentiates between detections and non-detections is not one of the plotted parameters: it is the azimuthal angle of the SMBH with respect to the galaxy center. 
Since the maximum of the arrival time surface, $\Psi_{\rm Fermat}$, given the large external shear and, and the positions of the sources in the lens macro-models, is significantly displaced from the center of the galaxy, it is the azimuthal angle of the SMBH, rather than its radial offset distance, that determines whether that SMBH will produce extra images. Therefore, RXJ0911+0551 does not provide constraints on $\Delta r$ significantly different from VODKA regardless of a central image detection or non-detection.

The non-detection posterior PDF for $r_c$ has a mode and 95\% credible interval of 0.485 kpc and $0.084 < r_c < 0.868$ kpc, respectively. The mode is very similar to that of the detection posterior distribution, however, the credible interval is larger in this case as a non-detection does not rule out small $r_c$.

For the unlikely case of a detection in RXJ0911+0551, we assume the central image is detected with the same properties as the ones in our main sample, with $f_o$ equivalent to 9 magnitudes fainter than the estimated brightest quad image, $\vec r_o$ at the center of the lens macro-model, $\sigma_r$ defined by the HST astrometric precision, and $\sigma_f = 0.4 f_o$. The detection posterior PDF for $\Delta r$ is very similar to the VODKA prior, with mode and 95\% credible interval of 0.184 kpc and $0.022 < \Delta r < 0.389$ kpc, respectively. A detection of the central image in RXJ0911+0551 does not improve SMBH offset constraints from VODKA. For core size $r_c$ and $b$, a detection provides slightly tighter constraints. The prior distribution of $r_c$ is skewed to larger values than the rest of the QSO samples. Likewise, for a detection, the posterior PDF has a mode of 0.310 kpc and a 95\% credible interval of  of $0.119 < r_c < 0.641$. A central image detection in this QSO would lead to a tighter constraint on $r_c$. It is worth noting that there is no measured velocity dispersion for RXJ0911+0551, and we have adopted a $\sigma$ measurement equal to the average of the measured $\sigma$ in the main sample. The constraints on $b$ with this QSO, therefore, could change significantly with a velocity dispersion measurement.

For both $b$ and $r_c$, we see that the detection posterior PDFs yield slightly tighter constraints than the prior, while the non-detection case does not. Furthermore, both cases have very similar estimates, something also observed with $\Delta r$. In general, we see that RXJ0911+0551 does not provide any strong constraints on $\Delta r$, $r_c$, and $b$. Likewise, a detection or non-detection does not significantly improve the posterior PDFs. We summarize our constraints for RXJ0911+0551 in Figure \ref{fig:RXJ0911pdfs}.

Given that the wide radial distribution of images in this lens was the primary cause for the large external shear, which lead to the lack of constraints on $\Delta r$, we can tentatively conclude that quads with a very wide radial image distribution and/or large external shear are unlikely to provide any constraints on $\Delta r$, $r_c$, and $b$, regardless of central image detection or non-detection.


\bsp	
\label{lastpage}
\end{document}